\documentclass[%
 preprint, 
 amsmath,amssymb,
 aps, physrev,
]{revtex4-2}
\UseRawInputEncoding
\usepackage[utf8]{inputenc}
\usepackage{graphicx}
\usepackage{dcolumn}
\usepackage{bm}
\usepackage{textgreek}
\usepackage{mathtools} 
\usepackage{color} 
\usepackage{hyperref}

\begin{document}

\preprint{APS/123-QED}
\title{\textbf{Direct experimental measurement of many-body hydrodynamic interactions with optical tweezers}}

\author{Dae Yeon Kim}
\altaffiliation{These authors contributed equally to this work.}
\affiliation{Department of Chemical Engineering, Stanford University, Stanford, CA, USA \\}

\author{Sachit G. Nagella}
\altaffiliation{These authors contributed equally to this work.}
\affiliation{Department of Chemical Engineering, University of California, Santa Barbara, Santa Barbara, CA, USA \\}
\affiliation{Department of Chemical Engineering, Stanford University, Stanford, CA, USA \\}

\author{Kyu Hwan Choi}
\affiliation{Department of Chemical Engineering, University of California, Santa Barbara, Santa Barbara, CA, USA \\}
\affiliation{Department of Chemical Engineering, Stanford University, Stanford, CA, USA \\}

\author{Sho C. Takatori}
\affiliation{Department of Chemical Engineering, Stanford University, Stanford, CA, USA}
\thanks{Corresponding author}
\email{stakatori@stanford.edu}
\date{\today}

\begin{abstract}
Many-body hydrodynamic interactions (HIs) play an important role in the dynamics of fluid suspensions. 
While many-body HIs have been studied extensively using particle simulations, there is a dearth of experimental frameworks with which to quantify fluid-mediated multi-body interactions. 
To address this, we design an experimental method that utilizes optical laser tweezers for quantifying fluid-mediated colloidal interactions with exquisite precision and control.
By inducing translation-rotation hydrodynamic coupling between trapped fluorescently-labeled colloids, we obtain a direct reporter of few- to many-body HIs experimentally.
We leverage the torque-free nature of laser tweezers to enable sensitive measurements of signals between trapped colloids.
First, we measure the pair HI between a stationary tracer probe and a translating particle as a function of their separation distance.
We discover that our technique can precisely quantify distant fluid disturbances that are generated by $\sim$2 pN of hydrodynamic force at 12 particle radii of separation.
To study the effect of many-body HIs, we measure the rotational mobility of a probe in a three-particle setup and in a model material, a two-dimensional hexagonally-close-packed lattice, that undergoes oscillatory strain.
Respectively, we discover that the probe's rotation can reverse in certain three-body configurations, and we find that rotational mobility in the crystalline array is strongly attenuated by particle rigidity.
Experimental measurements are corroborated by microhydrodynamic theory and Stokesian Dynamics simulations with excellent agreement, highlighting our ability to measure accurately many-body HIs.
The robustness of our experimental methodology enables us to extend our theoretical framework to manipulate colloidal-scale fluid flows.
With experimental validation, we compute the required trajectory of a moving particle to induce a desired angular velocity of a probe.
These findings underscore the potential of using colloidal agents to generate flows for directed, non-equilibrium assembly.

\end{abstract}


\maketitle


\section{\label{sec:intro}Introduction}
Owing to their small size, the dynamics of colloidal-scale particulates operates at low-Reynolds numbers.
Under an applied force, each particle's motion generates a disturbance flow in the surrounding fluid that decays algebraically like $\mathcal{O}(r^{-1})$, where $r$ is the radial distance.
This long-ranged flow field entrains neighbors, which respond by generating weaker disturbance fields. 
The propagation of forces between particles through the surrounding fluid, or hydrodynamic interactions (HIs), has been the subject of extensive study, in part due to the important role it plays in a number of contexts.
A notable example is sedimentation of colloidal particles under gravity.
Particles are caught in each other's flows, resulting in chaotic trajectories \cite{Guazzelli2011-gj} that trigger large-scale collective motion \cite{shen2020hydrodynamic}.
Theoretical analysis of the velocity fluctuations yields divergent scaling with system size \cite{Caflisch1985-td} in the absence of confining boundaries \cite{Brenner1999-zu, Nguyen2004-gx}, but mean sedimentation rates are attainable using renormalization methods \cite{Batchelor1972-it, Hinch1977-bg, Phillips2020-ua}.
In material science, the structure and stability of colloidal gels are highly dependent on fluid-mediated coupling between constituents \cite{varga2016hydrodynamic, Royall2015-tz, torre2023hydrodynamic}.
In biology, the beating of anchored lung cilia generates mucosal flows that expel colloidal-scale contaminants \cite{Smith2008}.
Similarly, we have recently shown that moving boundaries present as ``dynamic'' obstacles to particle transport, generating local flows which determine macroscopic diffusion \cite{nagella2023colloidal}.
Motile microorganisms experience no applied forces, but instead they exert force-dipoles \cite{Ishikawa2006-mq, Stone1996-av} on the surrounding fluid that decay as $\mathcal{O}(r^{-2})$, also facilitating collective motion \cite{Lushi2014-de, Ge2025-bv}.

Furthermore, advancements in optical imaging and manipulation using laser tweezers have granted greater command over the microscopic domain, with ever-increasing ability to probe nano- to micro-meter lengthscales with femto- to pico-Newton level forces \cite{Grier2003-be, xu2023dynamic, Zhang2022-yr, Volpe2023-wp}.
In particular, we believe that controlling local fluid flows, by selectively actuating individual bodies, will enable precise control over the local microstructure.
To these ends, we seek fundamental understanding of the solvent-mediated hydrodynamic forces between suspended particles.
This is further complicated by the fact that fluid-mediated interactions are not simply described by a pairwise summation of two-body interactions, as is usually assumed for conservative pair-potentials. 
Instead, the long-ranged nature of the disturbance flows generated by moving particles gives rise to non-negligible multi-body interactions.
Moreover, despite the wealth of theoretical techniques \cite{Mazur1982-of, Kynch1959-jj, Brady1988-ed}, there is a dearth of experimental frameworks with which to quantify precisely such fluid-mediated many-body interactions.

Studies of fundamental hydrodynamic interactions (e.g., among two or three particles) using microscopy have largely focused on the induced translational motion due to the flows produced by other particles' translational motion \cite{Crocker1997-io, Lele2011-uw, Lee2023-lf, Dufresne2000-ts}. 
Instead, we are concerned with probing the coupling between these flows and particle orientation.
Prior work has measured the hydrodynamic coupling between linear and rotational velocities by analyzing the equilibrium fluctuations in the particle positions and orientations and computing their correlations  \cite{Martin2006-aj, Zheng2022-tt}. 
We wish to measure the translation-rotation hydrodynamic coupling directly.
Even with existing high-resolution imaging techniques, measuring the angular displacement of a probe, specifically due to thermal forces on a neighboring particle, is obfuscated by torque fluctuations on the probe itself.
Therefore, we still lack a robust experimental framework to measure many-body hydrodynamic interactions directly and precisely.
To address these challenges, we use anistropic fluourescence labeling and optical tweezer manipulation to measure the rotational velocity of a tracer probe due to the disturbance flows generated by surrounding colloids.
We focus a laser on a particle's center-of-mass, applying a harmonic trapping force with a user-specified stiffness.
We use a sufficiently large trap stiffness to minimize the deviations in the particle position from the imposed laser position.
Note that this also obscures measurements of any translational displacements from other forces.
In doing so, we translate particles with velocities that are predetermined by the laser trajectories.
However, we emphasize that the laser does not impose an external torque on the trapped colloid, and the colloid is free to rotate.
Therefore, any induced hydrodynamic torque on the tracer probe will result in a direct signal of fluid disturbances that is not attenuated by external torques, enabling precise and sensitive measurements of angular displacements.
We can impose translational motion on the surrounding particles and directly measure the induced rotational velocity of a tracer probe to report the translation-rotation hydrodynamic coupling.

We use optical tweezers to translate particles with specified velocities and measure the induced rotational velocity of a stationary tracer probe.
The theoretical problem of predicting the probe's rotational velocity has features of both mobility and resistance perspectives. 
In a mobility formulation, one is given the hydrodynamic forces on the particles and determines the resulting velocities, and vice-versa for the resistances.
Here, the particles experience resistive drag forces to their imposed linear motion.
The propagation of these forces through the solvent exerts a torque on the probe to produce rotational motion.
For a pair of particles, we develop an analytical expression for the induced rotational velocity of the probe using the method of reflections, pioneered by Smoluchowski \cite{Smoluchowski1911-du} and adopted by several workers with successful results \cite{Kynch1959-jj, Beenakker1982-vs, Mazur1982-of, Kim1985-mt, Happel2012-fp}, following a matrix inversion process \cite{Ichiki2001-yx}. 
Similarly, in a multiparticle setup, we resolve the effect of many-body hydrodynamic interactions using the Stokesian Dynamics paradigm \cite{Durlofsky1987-oc, Brady1988-ed, Brady1988-yd, Sierou2001-ks, Banchio2003-ar}.
Our computational approach is inspired by that of Fiore and Swan \cite{Fiore2019-or}, who have reformulated the Stokesian Dynamics method for large-scale, soft matter simulations.
They show that the details of mixing mobility and resistance problems are easily accounted through a saddle point formulation, where the forces and velocities may be determined simultaneously.
Overall, we found excellent agreement between the experimental data and our computations.

We present a combined effort of optical tweezer experiments, microhydrodynamic theory, and Stokesian Dynamics simulations to study the impact of local flows on particle orientation. 
By labeling the particle hemispheres with a fluorescent marker, we track orientation displacements due to translation-rotation hydrodynamic coupling.
We demonstrate that our methodology faithfully captures the two-body hydrodynamic interaction between a rotating probe and a nearby, translating particle.
Upon introducing a third particle, our results become highly dependent on the particular three-particle configuration.
In some cases, the probe's rotation reverses because of a three-body interaction that arises from induced ``backflows''.
The role of many-body HIs is further demonstrated in a model material, consisting of a hexagonally-closed-packed (HCP) lattice of particles.
We apply an oscillatory strain on the lattice and measure the dependence of the center sphere's rotation rate on local packing.
Our results illustrate the exquisite sensitivity and precision with which we can report many-body HIs by measuring particle rotations.
Finally, as a demonstration of directing particle rotation by optical manipulation of local fluid flows, we return to the two-particle setup and consider the inverse problem: specifying a desired rotation rate of the probe and determining the trajectory of the moving particle.

The rest of this paper is organized as follows: in Sec.~\ref{sec:methods} we describe our experimental and theoretical methods.
In Sec.~\ref{sec:results} we discuss our experimental results and theoretical predictions.
Lastly, we offer closing remarks and provide suggestions for future work in Sec.~\ref{sec:conclusion}.
\section{\label{sec:methods}Materials and Methods}
\subsection{Experimental Setup}

\begin{figure}[h!]
\centering
\includegraphics[width=\columnwidth]{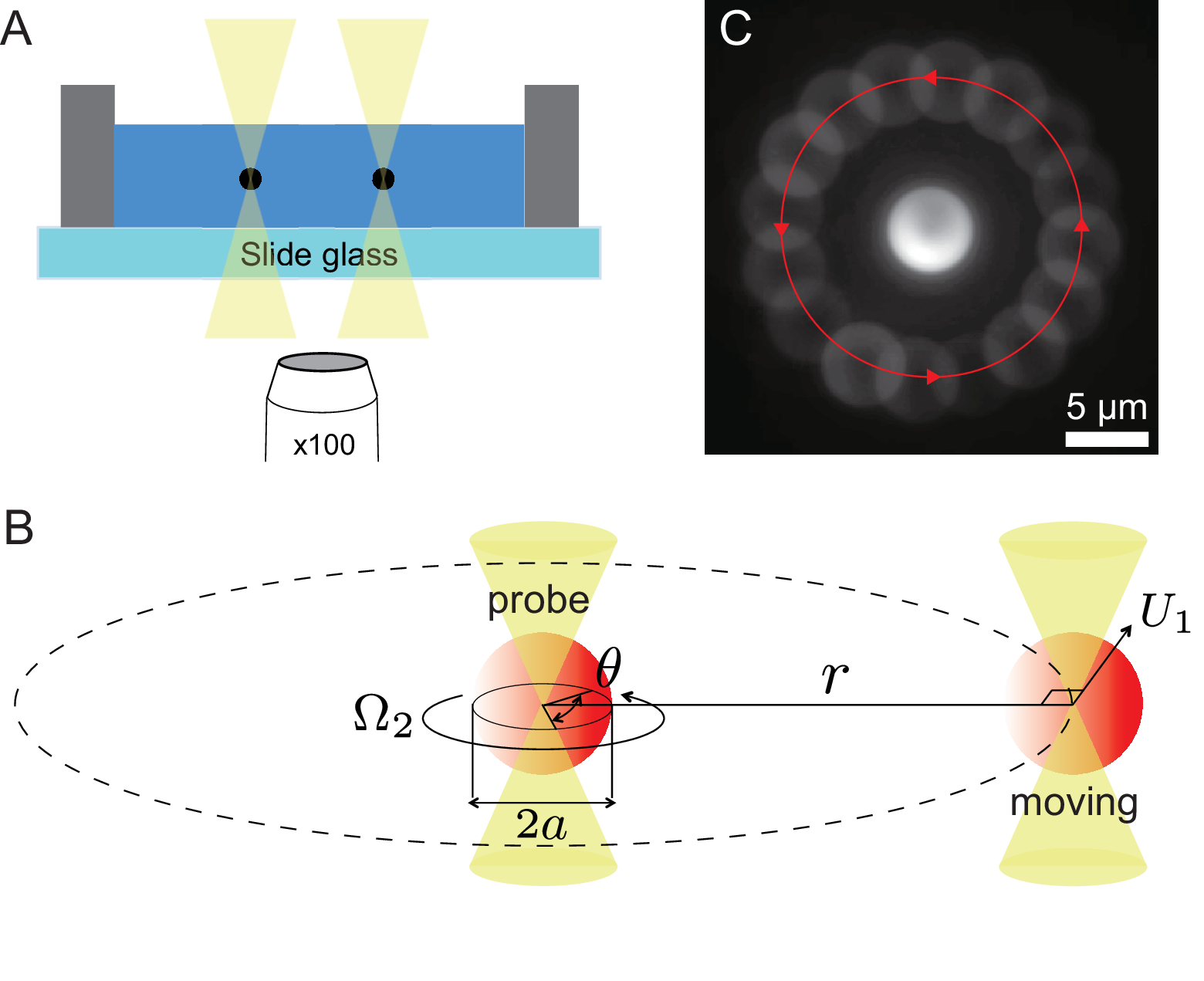}
\caption{\label{fig:fig1} Experimental setup of measuring the induced rotational motion a colloidal probe due to a nearby, translating particle. (A) Schematic of optical laser tweezers and trapped particles in solution. (B) We used strong traps to hold a pair of particles of radius $a = 2.5$ \textmu m at a fixed pair separation distance, $r$. Colloidal surfaces are fluorescently labeled to track the particle orientations. We imposed a circular trajectory on the moving particle with speed $U_1 = 35-200$ \textmu m/s. In response, the probe rotated with velocity $\Omega_2 = a\skew{4}{\dot}{\theta}$, where $\theta$ was the measured polar angle and $\skew{4}{\dot}{\theta}$ was the angular velocity. (C) Superimposed image of many frames of a translating outer particle. See Supplemental Movie 2 and 3 for a video of the experiment.}
\end{figure}

\subsubsection{Colloidal particles with hemispherical coating of fluorescent dye}
In this study, it was crucial to measure the time-dependent orientation of the particles to observe the hydrodynamic coupling between their rotations and the translational motion of their neighbors. 
To visualize the orientation, we applied a hemispherical fluorescent dye coating to spherical particles and used a suspension of these half-coated polymethyl methacrylate (PMMA) particles as the working fluid. 
Monodisperse amino PMMA latex (5\% w/v, Abvigen Inc, USA) with a size of 5(±0.05) \textmu m was utilized, and the particle concentration of the suspension for the experiments was set at 0.2 vol\%. 

The hemispherical coating of the PMMA particles was achieved by applying the gel trapping technique (GTT) \cite{paunov2003novel}. 
First, a 2 wt \% gellan gum aqueous solution (Phytagel, Sigma-Aldrich, USA) was poured into a Petri dish and n-Decane (Sigma-Aldrich, USA) was added to create an oil-water interface. 
A solution, containing washed amino PMMA particles mixed with isopropyl alcohol (IPA), was carefully introduced to the interface using a micropipette, thereby positioning the particles at the interface. 
Here, IPA acted as a spreading solvent. 
The system was then cooled at room temperature for approximately one hour, allowing gelation of the gellan gum solution beneath the interface. 
After n-Decane was removed, a poly(dimethylsiloxane) (PDMS) silicone elastomer (Sylgard 184, Dow, USA) was poured over the gelled solution and left to cure for two days. 
Upon curing, the PDMS was carefully peeled off from the gellan gum, capturing the PMMA particles that were partially embedded on one side. 
To fluorescently label the particles, the amine groups on the surface of the particle reacted with Alexa Fluor 647 NHS Ester (Invitrogen, USA) by NHS ester-amine coupling. 
The dyed particles were then collected using water-soluble polyvinyl alcohol (PVA) tape dissolved in deionized (DI) water, and centrifuged to isolate the particles. 
The isolated particles were washed multiple times with DI water and subsequently diluted for experimental use.
This preparation method ensured a consistent hemispherical fluorescent coating on PMMA particles as shown in the center particle of Fig.\ref{fig:fig1}C, allowing accurate orientation measurements during experiments. 

\subsubsection{Optical trapping and observation setup}
Colloidal particles were manipulated using optical laser tweezers (Tweez 305, Aresis Ltd, Slovenia) equipped with a continuous infrared laser (wavelength: 1064 nm, maximum power: 5 W). 
The system was integrated with an inverted fluorescence microscope (Ti2-Eclipse, Nikon, Japan) fitted with an oil immersion Apo 100× objective lens (Numerical Aperture (NA): 1.45, Nikon, Japan).
For excitation, a Lumencor SpectraX Multi-Line LED Light Source (Lumencor, Inc., USA) was used at a wavelength of 647 nm. 
Fluorescent emissions were spectrally filtered using a multi-band emission filter (680/42; Semrock, IDEX Health \& Science, USA). 

The measurement schematic is shown in Fig.\ref{fig:fig1}A. Two particles were trapped at a fixed pair separation distance, $r$. 
They were positioned far from the walls, air-water interface, and the bottom of the observation cell to avoid confinement effects. 
Additionally, the experiments were conducted under dilute conditions where the presence of other particles, besides the trapped ones, could be neglected. 
Subsequently, as depicted in Fig.\ref{fig:fig1}B, the moving particle was driven along a circular trajectory at a constant speed, $U_1$, while the rotation of the centrally trapped probe particle was recorded using a CMOS camera (Photometrics Prime 95B, Teledyne Photometrics, USA) at 100 frames per second (fps). 
We varied $U_1$ from 35 \textmu m/s to 200 \textmu m/s, and $r$ from 5 \textmu m to 30 \textmu m. 
At these experimental conditions, the Reynolds number (\( \text{Re}_p = \rho U_1 d/\eta \), where \( \rho \) is the fluid density, \( U_1 \) is the speed of the moving particle, \( d \) is the particle diameter, and \( \eta \) is the shear viscosity of the medium) was $\mathcal{O}(10^{-4}-10^{-3})$. 
The Stokes number (\( \text{St} = \frac{\rho_p d^2}{18 \eta} \frac{U_1}{L} \), the ratio of the characteristic time of a particle to the characteristic time of the flow, where \( \rho_p \) is the particle density and \( L \) is the characteristic length, which, in this case, corresponds to the particle diameter of 5 \textmu m) was $\mathcal{O}(10^{-5})$. 
Additionally, the P\'eclet number of the moving particle (\( \text{Pe}_p = U_1L/D \), the ratio of the diffusion timescale to the advection timescale, where \( D \) is the Stokes-Einstein-Sutherland diffusivity) was $\mathcal{O}(10^{3}-10^{4})$. 
The high Peclet numbers (\( \text{Pe}_p \gg 1 \)) suggest that the driving forces we apply overwhelm thermal fluctuations. 
Our experiments are conducted at low Reynolds numbers, providing a robust basis for quantitative analysis of hydrodynamic interactions throughout this work. 
A superposition of many frames of a timelapse movie is shown in Fig.\ref{fig:fig1}C. Further details on the optical tweezer and the observation cell are found in the study by Xu et al.~\cite{xu2023dynamic}.

\subsubsection{Measurement of rotational angle}
The polar angle of the probe particle, $\theta$, was determined by tracking the time-dependent orientation of the hemispherically fluorescent-coated particle.
The fluorescently coated hemisphere facilitated clear visualization of the particle’s orientation under fluorescence imaging.
Using an in-house MATLAB image analysis script, the bright fluorescent region was identified in each frame of the recorded images, and the particle’s orientation angle was extracted with respect to a fixed reference axis.
To validate our measurements, we analyzed the Brownian motion of a single trapped particle by measuring the mean square angular displacement (MSAD). Our results showed excellent agreement with the 2D Smoluchowski-Einstein-Sutherland (SES) equation (see Sec.~\ref{sec:appendix_exp_rotational diffusivity}). Notably, out-of-plane rotation was seldom observed, likely due to the vertical alignment induced by the laser tweezer on the particle’s slight but non-negligible asymmetry and surface roughness. This effect is consistent with previous observations for rod-like particles, such as \textit{Escherichia coli} and other elongated microorganisms, where the laser trap pins the vertical orientation of the body and suppresses rotational motion out of the focal plane \cite{s2016optical, zhang2019manipulating, yadav2020optimal}. While our experiments produce 2D rotational motion, the problem remains fully three-dimensional in nature.
The experimentally obtained diffusivity exhibited excellent agreement with the theoretical predictions, confirming the accuracy of our methodology. A detailed explanation of this validation can be found in Sec.~\ref{sec:appendix_exp_rotational diffusivity}.

\subsection{Microhydrodynamic Theory}
\subsubsection{Far-field hydrodynamics}
The translational motion of the moving particle disturbs the surrounding fluid, exerting a hydrodynamic torque that rotates the probe about its center-of-mass (Fig.~\ref{fig:fig1}C). 
We seek an analytical expression for the probe's rotational velocity, $\bm{\Omega}_2$, due to the imposed translation on the neighboring sphere, $\bm{U}_1$.

We find the spatially-dependent velocity field, $\bm{u}({\bm{x}})$, in the solvent phase using the equations of Stokes flow,
\begin{gather}
    \eta\nabla^2\bm{u} - \nabla p = \bm{f}, \label{eq:stokes} \\ 
    \nabla \cdot \bm{u} = 0, \label{eq:continuity}
\end{gather}
subject to a body force, $\bm{f}$. 
The fluid pressure, $p$, enforces incompressibility of the fluid, Eqn.~\ref{eq:continuity}.
The velocity field satisfies the no-slip condition at the surface of the spheres, corresponding to rigid body motion.

In the experiments, the two spheres are suspended in a very dilute solution and are positioned far from the boundaries. 
Thus, we may safely ignore the effect of the confining boundaries and assume that the disturbance field decays ``infinitely away'' from the particles. 

As a first approximation, an immersed, moving particle applies a point-force on the fluid, $\bm{f} = \bm{F}_0\delta(\bm{x})$, with amplitude $\bm{F}_0$.
For an unbounded domain, the Green's function that solves Eqs.~\ref{eq:stokes} and \ref{eq:continuity} is the Oseen tensor, 
\begin{equation}\label{oseen}
    \bm{J}(\bm{x}) = \frac{1}{r}\left( \bm{I} + \hat{\bm{x}}\hat{\bm{x}} \right), 
\end{equation}
and the velocity field is $\bm{u}(\bm{x}) = \frac{1}{8\pi \eta} \bm{J}(\bm{x})\cdot \bm{F}_0$.
The radial distance from the origin is $r = \left( \bm{x} \cdot \bm{x} \right)^{1/2}$, $\hat{\bm{x}} = \bm{x}/r$, and $\bm{I}$ is the identity tensor.
The long-ranged $\mathcal{O}(r^{-1})$ decay is the crux of hydrodynamic phenomena involving many particles. 
In a suspension, each particle's motion disturbs the solvent, generating local flows that entrain both close and distant neighbors.
Also, the lack of time-dependence in Eqn.~\ref{eq:stokes} implies that the instantaneous spatial configuration of particles and their velocities determine the disturbance flow, which quasi-statically evolves with the particles' dynamics.
Several theoretical or computational approaches have been developed to tackle problems involving hydrodynamic interactions \cite{Tornberg2008-mt, Ladd1988-rs, Ganatos1978-he, Mo1994-ju, Pozrikidis1992-nj}, and here we review one \cite{Fiore2019-or}.

To account for the finite size of the suspended particle, we take its surface to be a tessellation of infinitely many point-forces on the surrounding fluid. 
By linearity of the Stokes equations, the resulting velocity field is a superposition of the corresponding point-force solutions, converging to an integral over the surface of the particle \cite{Ladyzhenskaia1963-ll},
\begin{equation}\label{eq:vel_int_soln}
    \bm{u}(\bm{x}) = \frac{1}{8\pi\eta}\oint_{S_{\beta}}{\bm{J}(\bm{x}-\bm{y})\cdot\bm{f}_\beta(\bm{y}})\text{d}S(\bm{y}).
\end{equation}
We define the traction vector, $\bm{f}_\beta = \bm{\sigma}\cdot \bm{n}_\beta$, as the projection of the fluid stress tensor, $\bm{\sigma}$, onto the normal vector, $\bm{n}$, pointing out of the particle surface. 
Physically, the Green's function propagates the distribution of forces on the particle surface, $\bm{f}_\beta$, into the solvent and generates flow.

Supposing a nearby particle, $\alpha$, is immersed in the flow field produced by particle $\beta$, Eqn.~\ref{eq:vel_int_soln}, we wish to compute its motion.
For widely-separated spheres, we expand the traction vector into its first few multipole moments, 
\begin{equation} \label{eq:multipole_exp}
    \bm{f}_\beta(\bm{y}) \approx \frac{1}{4\pi a^2}\bm{F}^{\text{H}}_\beta + \frac{3}{4\pi a^3}\bm{n}(\bm{y}) \cdot \bm{C}^{\text{H}}_\beta + ... \ , 
\end{equation}
including the hydrodynamic force (0th moment), 
\begin{equation} \label{force}
    \bm{F}^{\text{H}}_\beta = \oint_{S_{\beta}} \bm{f}_\beta(\bm{y}) \text{d}S, 
\end{equation}
and truncating at the level of the force-dipole (1st moment),
\begin{equation} \label{couplet}
    \bm{C}^{\text{H}}_\beta = \oint_{S_{\beta}} \left[ \bm{y}\bm{f}(\bm{y}) - \frac{1}{3}\bm{I}\left(\bm{y}\cdot\bm{f}(\bm{y})\right) \right] \text{d}S . 
\end{equation}
The symmetric and anti-symmetric components of $\bm{C}^{\text{H}}_\beta$ are the particle stresslet \cite{Batchelor1970-kb}, $\bm{S}^{\text{H}}_\beta = \frac{1}{2}\oint_{S_\beta}\bm{y}\bm{f}_{\beta} + \bm{f}_\beta\bm{y} - \frac{2}{3}\bm{I}\left(\bm{y}\cdot\bm{f}_\beta\right)\text{d}S$, and the hydrodynamic torque, $\bm{L}^{\text{H}}_\beta = \oint_{S_\beta} \bm{y}\times\bm{f}_\beta\text{d}S$, respectively.
From the disturbance field, we determine the particle motion using Fax\'en's laws \cite{Batchelor1972-it, Fiore2018-ax, Fiore2019-or, Batchelor1972-hq}, 
\begin{align}
    \label{faxen_int_vel}
    \bm{U}_{\alpha} = &\frac{1}{4\pi a^2}\oint_{S_{\alpha}} \bm{u}(\bm{x}) \text{d}S, \\
    \label{faxen_int_def}
    \bm{D}_{\alpha} = &\frac{3}{4\pi a^3} \oint_{S_{\alpha}}\bm{u}(\bm{x})\bm{n}_\alpha\text{d}S ,
\end{align}
in which $\bm{U}_{\alpha}$ is the translational velocity of particle $\alpha$ and $\bm{D}_{\alpha}$ is its deformation tensor, whose symmetric and anti-symmetric parts are the particle rate-of-strain, $\bm{E}_\alpha = \frac{3}{8\pi a^3}\oint_{S_\alpha} \bm{u}\bm{n}_\alpha + \bm{n}_\alpha\bm{u}\text{d}S$, and rotational velocity, $\bm{\Omega}_\alpha = \frac{3}{4\pi a^2}\oint_{S_\alpha} \bm{u}\times \bm{n}_{\alpha} \text{d}S $, respectively.
Substituting the truncated multipole expansion, Eqn.~\ref{eq:multipole_exp}, into Eqn.~\ref{eq:vel_int_soln} and evaluating Fax\'en's laws yields a linear relationship between the induced motion of particle $\alpha$ and the force moments on particle $\beta$, 
\begin{equation}
    \begin{pmatrix}
        \bm{U}_{\alpha} \\
        \bm{D}_{\alpha}
    \end{pmatrix}
    =
    \begin{pmatrix}
        \bm{M}^{\text{UF}}_{\alpha \beta} & \bm{M}^{\text{UC}}_{\alpha \beta} \\ 
        \bm{M}^{\text{DF}}_{\alpha \beta} &
        \bm{M}^{\text{DC}}_{\alpha \beta}
    \end{pmatrix}
    \begin{pmatrix}
        \bm{F}^{\text{H}}_{\beta} \\
        \bm{C}^{\text{H}}_{\beta}
    \end{pmatrix} .
\end{equation}
The mobility tensors describing pairwise hydrodynamic interactions depend on the vector of separation that connects the pair, $\bm{r}_{\alpha \beta} = \bm{x}_{\alpha} - \bm{x}_{\beta}$, and are expressed as integrals of the Green's function over the spherical surfaces,  
\begin{eqnarray}
    \label{eq:muf_int}
    \bm{M}_{\alpha \beta}^{\text{UF}} = \frac{1}{4\pi a^2}\oint_{S_{\alpha}} \text{d}S \ \frac{1}{4\pi a^2}\oint_{S_{\beta}} \bm{J}(\bm{r}_{\alpha \beta})\text{d}S, \\
    \label{eq:muc_int}
    \bm{M}_{\alpha \beta}^{\text{UC}} = \frac{1}{4\pi a^2}\oint_{S_{\alpha}} \text{d}S \ \frac{3}{4\pi a^3}\oint_{S_{\beta}} \bm{J}(\bm{r}_{\alpha \beta}) \bm{n}_{\beta} \text{d}{S}, \\
    \label{eq:mdf_int}
    \bm{M}_{\alpha \beta}^{\text{DF}} = \frac{3}{4\pi a^3}\oint_{S_{\alpha}} \bm{n}_{\alpha}\text{d}S \ \frac{1}{4\pi a^2}\oint_{S_{\beta}} \bm{J}(\bm{r}_{\alpha \beta}) \text{d}{S}, \\
    \label{eq:mdc_int}
    \bm{M}_{\alpha \beta}^{\text{DC}} = \frac{3}{4\pi a^3} \oint_{S_{\alpha}} \bm{n}_{\alpha} \text{d}S \ \frac{3}{4\pi a^3}\oint_{S_{\beta}} \bm{J}(\bm{r}_{\alpha \beta}) \bm{n}_{\beta} \text{d}{S} .
\end{eqnarray}
One can recover the well-known differential forms of the Fax\'en and mobility relations by expanding the Green's function around each particle position and applying the harmonic properties of the homogeneous Stokes equations.
Analytical expressions for Eqs.~\ref{eq:muf_int}-\ref{eq:mdc_int} that we use in this work are tabulated elsewhere \cite{Kim1991-cs, Durlofsky1987-oc}.
\subsubsection{The induced rotation rate of a probe due to a translating sphere of equal size}
Linearity of Stokes flow implies that the rotational motion of the probe is linearly proportional to the imposed translational velocity of the moving particle. 
To deduce this relationship, we construct a symmetric matrix of pair mobility tensors that relate the particle kinematics to the multipole moments of the hydrodynamic traction \cite{Happel2012-fp, Kim1991-cs}.
For now, accounting for the particle forces and torques is sufficient.
In the following section, we consider interactions among many particles, and the particle stresslets become important.
Labeling the moving particle as ``1'' and the probe ``2'', the linear relationship between particle-pair kinematics and corresponding multipole moments is 
\begin{eqnarray}\label{eq:mobility_relation}
    \begin{pmatrix}
        \bm{U}_1 \\
        \bm{U}_2 \\
        \bm{\Omega}_1 \\
        \bm{\Omega}_2 \\
    \end{pmatrix}
    = 
    \begin{pmatrix}
        \bm{M}^{\text{UF}}_{11} & \bm{M}^{\text{UF}}_{12} & \bm{M}^{\text{UL}}_{11} & \bm{M}^{\text{UL}}_{12} \\
        \bm{M}^{\text{UF}}_{21} & \bm{M}^{\text{UF}}_{22} & \bm{M}^{\text{UL}}_{21} & \bm{M}^{\text{UL}}_{22} \\
        \bm{M}^{\Omega \text{F}}_{11} & \bm{M}^{\Omega \text{F}}_{12} & \bm{M}^{\Omega \text{L}}_{11} & \bm{M}^{\Omega \text{L}}_{12} \\
        \bm{M}^{\Omega \text{F}}_{21} & \bm{M}^{\Omega \text{F}}_{22} & \bm{M}^{\Omega \text{L}}_{21} & \bm{M}^{\Omega \text{L}}_{22}
    \end{pmatrix}
    \begin{pmatrix}
        \bm{F}^{\text{H}}_1 \\
        \bm{F}^{\text{H}}_2 \\
        \bm{L}^{\text{H}}_1 \\
        \bm{L}^{\text{H}}_2 
    \end{pmatrix} ,
\end{eqnarray}
succinctly expressed as $\mathcal{U} = \mathcal{M} \cdot \mathcal{F}$. The symmetric $6N \times 6N$ matrix, $\mathcal{M}$, which couples the $N$ particles' motion to the hydrodynamic forces and torques, is the grand mobility matrix \cite{Happel2012-fp}.
Importantly, we ensure that it remains symmetric and positive-definite, a consequence of the dissipative nature of Stokes flow, throughout all computations.
To determine the relationship between $\bm{\Omega}_2$ and $\bm{U}_1$, we couple the mobility relation, Eqn.~\ref{eq:mobility_relation}, to the particle force and torque balances.
In the overdamped limit, these are 
\begin{eqnarray} 
    \bm{F}^{\text{H}}_\alpha + \bm{F}^{\text{ext}}_\alpha &= \bm{0} , \label{fbalance} \\
    \bm{L}^{\text{H}}_\alpha + \bm{L}^{\text{ext}}_\alpha &= \bm{0} \label{lbalance},
\end{eqnarray}
ignoring thermal fluctuations.
The hydrodynamic forces and torques are balanced by externally applied fields. 
In particular, the laser tweezers exert a force to maintain prescribed translational velocities, $\bm{U}_\alpha$, while the particles rotate freely (i.e., $\bm{L}^{\text{ext}}_\alpha = \bm{0}$).
From these constraints, we simplify Eqn.~\ref{eq:mobility_relation}, 

\begin{eqnarray}
    \begin{pmatrix}
        \hat{\textbf{U}} \\
        \hat{\bm{\Omega}}
    \end{pmatrix}
    = 
    \begin{pmatrix}
        \textbf{M}^{\text{UF}} & \textbf{M}^{\text{UL}} \\
        \textbf{M}^{\Omega \text{F}} & \textbf{M}^{\Omega \text{L}}
    \end{pmatrix}
    \begin{pmatrix}
        \hat{\textbf{F}}^{\text{H}}
        \\
        \bm{0}
    \end{pmatrix} , 
\end{eqnarray}
where we use block matrices and stacked vectors for brevity.
Then, one can solve for the required hydrodynamic forces (equivalently, the externally applied forces) to maintain the imposed translational motion, and use these to determine the rotational velocities.
The result is 
\begin{eqnarray}
    \hat{\bm{\Omega}} = \textbf{M}^{\Omega \text{F}} \cdot \left( \textbf{M}^{\text{UF}} \right)^{-1} \cdot \hat{\textbf{U}} ,
\end{eqnarray}
and the rotational velocity of the probe is 
\begin{eqnarray}\label{eq:probe_moving_interaction}
    \bm{\Omega}_2 = y^{*}(r_{12}/a)\hat{\bm{r}}_{21}\times\bm{U}_1 .
\end{eqnarray}
The normalized pair-separation vector is $\hat{\bm{r}}_{\alpha \beta} = \bm{r}_{\alpha \beta}/r_{\alpha \beta}$ and  
\begin{eqnarray} \label{eq:far_field_interaction}
    y^{*}(s) =
\frac{3s}{-4 - 6s^2 + 8s^3} + \frac{3s}{4 + 6s^2 + 8s^3} 
\end{eqnarray}
is a dimensionless, scalar function describing the hydrodynamic interaction.
This result holds for widely separated spheres.
For closely touching spheres, we follow a similar approach using asymptotic expressions of the the pair hydrodynamic resistances developed by Jeffrey and Onishi \cite{Jeffrey1984-jk, Jeffrey1992-lc}. 
This is equivalent to carrying out the mulitpole expansion in Eqn.~\ref{eq:multipole_exp} for a single particle to an infinite number of moments, thereby resolving the interaction between nearly-touching particles.
Whereas the result for the far-field interaction is expressed in terms of the center-to-center distance, the near-field interaction is best described using the surface-to-surface distance, $\xi = (r_{12} - 2a)/a$.
Details of the calculation are in Sec.~\ref{sec:appendix_nf_interaction}. 
The result for the rotational velocity has the same mathematical structure as Eqn.~\ref{eq:probe_moving_interaction}, with the scalar function possessing the form
\begin{eqnarray} \label{eq:near_field_interaction}
    Y^{*}(\xi) = \prod_{m = 0}^{1}\left[ \sum\limits_{n = 0}^{3}h_{mn}(\xi)\ln(\xi)^n \right]^{2m-1}.
\end{eqnarray} 
The coefficients, $h_{mn}(\xi)$, are power series in $\xi$ of order $n$ (see Sec.~\ref{sec:appendix_nf_interaction}).

Equations \ref{eq:far_field_interaction} and \ref{eq:near_field_interaction} are the main results of this section, and in the following, we extend our analysis to develop a numerical scheme suitable for multiparticle systems.

\subsubsection{Multiparticle interactions}
Here, we extend the analysis of the previous section to handle computing interactions among many particles. 
Including the particle stresslets in the multipole expansion has been shown to resolve many-body effects by computing the scattering of the force-dipoles off of all neighbors \cite{Brady1988-ed, Brady1993-ru}.

Augmenting the mobility relation, Eqn.~\ref{eq:mobility_relation}, from the previous section with the particle rate-of-strain and stresslet, we have 
\begin{eqnarray} \label{eq:stress_augmented}
    \begin{pmatrix}
        \mathcal{M} & \textbf{M}^{\mathcal{U}\text{S}} \\
        \textbf{M}^{\text{E}\mathcal{F} } & \textbf{M}^{\text{ES}}
    \end{pmatrix}
    \begin{pmatrix}
        \mathcal{F} \\
        \hat{\textbf{S}}^{\text{H}}
    \end{pmatrix}
    = 
    \begin{pmatrix}
        \mathcal{U} \\
        \hat{\textbf{E}}
    \end{pmatrix}.
\end{eqnarray}
Rigid particles cannot deform, $\hat{\textbf{E}} = \bm{0}$, imposing a constraint that enables solving for the particle stresslets, 
\begin{eqnarray}\label{eq:stresslet_constraint}
    \hat{\textbf{S}}^{\text{H}} = -\left( \textbf{M}^{\text{ES}} \right)^{-1} \cdot \textbf{M}^{\text{E}\mathcal{F}}\cdot \mathcal{F}.
\end{eqnarray}
We obtain a modified mobility relation,
\begin{eqnarray} \label{eq:multi_mob_relation}
    \mathcal{U} = \mathcal{M}^{\dagger} \cdot \mathcal{F} ,
\end{eqnarray}
identifying a ``stresslet-constrained'' mobility matrix, $\mathcal{M}^{\dagger} = \mathcal{M} - \textbf{M}^{\mathcal{U}\text{S}} \cdot \left( \textbf{M}^{\text{ES}} \right)^{-1} \cdot \textbf{M}^{\text{E}\mathcal{F}} $ \cite{Fiore2018-ax, Brady1988-ed}.
The disturbance flows in the solvent cannot deform the immersed particles, hindering their motion. 

As before, we couple the mobility relation to the torque balance, 
\begin{eqnarray}
    \bm{L}^{\text{H}, \text{ff}}_\alpha + \bm{L}^{\text{H}, \text{nf}}_\alpha + \bm{L}^{\text{ext}}_\alpha = \bm{0},
\end{eqnarray}
splitting the hydrodynamic torque into far-field and near-field contributions.
The far-field hydrodynamic torques (and forces) are obtained by inverting the mobility relation, $\mathcal{F} = \left( \mathcal{M}^{\dag}\right)^{-1} \cdot \mathcal{U}$. 
Note that we do not explicitly perform this operation in the formulation presented here.
The near-field lubrication interactions contained in $\bm{L}^{\text{H,nf}}_\alpha$ are a pairwise summation of the known, pair resistance tensors \cite{Jeffrey1992-lc, Jeffrey1984-jk}.
The pair resistance matrices, 
\begin{equation}
\mathcal{R}^{\text{2B}} = 
\begin{bmatrix}
    \textbf{R}^{\text{2B,FU}} & \textbf{R}^{\text{2B,F}\Omega} \\
    \textbf{R}^{\text{2B,LU}} & \textbf{R}^{\text{2B,L}\Omega}
\end{bmatrix},
\end{equation}
are exact series expansions for the hydrodynamic resistance between two spheres. 
Therefore, $\mathcal{R}^{\text{2B}}$ already contains the two-particle contribution from $(\mathcal{M}^{\dag})^{-1}$.  
To avoid double counting the two-body far-field resistances, we define the near-field resistance matrix, 
\begin{equation} \label{eq:r2b_nf_define}
\mathcal{R}^{\text{2B,nf}} = 
\mathcal{R}^{\text{2B}} - \left( \mathcal{M}^{\text{2B},\dag}\right)^{-1},
\end{equation}
where $\left( \mathcal{M}^{\text{2B},\dag} \right)^{-1}$ is the inverse of the grand mobility matrix for two particles.
By spherical symmetry, $\mathcal{R}^{\text{2B}}$ and $\mathcal{R}^{\text{2B,nf}}$ may be expressed in terms of scalar functions that are the coefficients to orthonormal tensor products involving the pair-separation vectors. 
Equation \ref{eq:r2b_nf_define} only needs to be calculated once, and the result is stored for pairwise construction of the near-field multiparticle resistance matrix \cite{Jeffrey1992-lc, Townsend2023-nc}.

The near-field hydrodynamic torques are $\skew{-4}\hat{\textbf{L}}^{\text{H,nf}} = -\textbf{R}^{\text{L}\Omega,\text{nf}} \cdot \hat{\bm{\Omega}} -\textbf{R}^{\text{LU,nf}} \cdot \hat{\textbf{U}}$. 
Using $\textbf{R}^{\text{L}\Omega,\text{nf}}$ as an example, we construct the near-field multiparticle resistance matrices in the following manner, 
\begin{eqnarray}
    \textbf{R}^{\text{L}\Omega,\text{nf}} = 
    \begin{bmatrix}
        \sum\limits_{\beta \neq 1}^N \bm{R}^{\text{L}\Omega,\text{nf}}_{11}(\bm{r}_{1\beta}) & \bm{R}^{\text{L}\Omega,\text{nf}}_{12}(\bm{r}_{12}) & ... & \bm{R}^{\text{L}\Omega,\text{nf}}_{12}(\bm{r}_{1N}) \\
        \bm{R}^{\text{L}\Omega,\text{nf}}_{21}(\bm{r}_{21}) & \sum\limits_{\beta \neq 2}^N \bm{R}^{\text{L}\Omega,\text{nf}}_{11}(\bm{r}_{2\beta}) & &  \vdots \\
        \vdots & & \ddots & \\
        \bm{R}^{\text{L}\Omega,\text{nf}}_{21}(\bm{r}_{N1}) & ...& &\sum\limits_{\beta \neq N}^N \bm{R}^{\text{L}\Omega,\text{nf}}_{11}(\bm{r}_{N\beta})
    \end{bmatrix} .
\end{eqnarray}
$\textbf{R}^{\text{LU},\text{nf}}$ is assembled similarly.
The higher order multipole moments within $\mathcal{R}^{\text{2B,nf}}$ decay very quickly, so we set $\mathcal{R}^{\text{2B,nf}}(\bm{r}_{\alpha \beta}) = \bm{0}$ for interparticle distances that are greater than a specified cutoff radius, $r_{\alpha \beta} > r_{\text{lub}}$. 
We choose $r_{\text{lub}} = 4a$.
Together, the mobility relation and torque balance are
\begin{eqnarray}
    \label{eq:mobility_relation_saddle}
    \mathcal{M}^{\dagger} \cdot \mathcal{F} + \mathcal{B} \cdot \hat{\bm{\Omega}} = 
    \begin{pmatrix}
        \hat{\textbf{U}} \\
        \bm{0}
    \end{pmatrix}, \\ 
    \label{eq:torque_bal_saddle}
    -\mathcal{B}^{T} \cdot \mathcal{F} - \textbf{R}^{\text{L}\Omega, \text{nf}}\cdot \hat{\bm{\Omega}} = -\left(\skew{-4}\hat{\textbf{L}}^{\text{ext}} - \textbf{R}^{\text{LU}, \text{nf}}\cdot \hat{\textbf{U}} \right).
\end{eqnarray}
The projection operator, $\mathcal{B}$, elevates the rotational velocities to the set of generalized velocities, such that $\mathcal{B} \cdot \hat{\bm{\Omega}} = \left(\bm{0} \ \ -\hat{\bm{\Omega}} \right)^{T}$ .
Its transpose, $\mathcal{B}^T$, projects the generalized set of particle motions and force-moments, $\mathcal{U}  \ \text{and} \ \mathcal{F}$, onto the subspace of rotations and torques, as demonstrated by the first term of Eqn.~\ref{eq:torque_bal_saddle}. 
For widely-spaced particles, $\textbf{R}^{\text{L}\Omega,\text{nf}} = \bm{0}$ and $\textbf{R}^{\text{LU},\text{nf}} = \bm{0}$. 
Further, in the absence of externally applied torques, the torque balance reduces to $-\mathcal{B}^{T} \cdot \mathcal{F} = \skew{-4}\hat{\textbf{L}}^{\text{H}, \text{ff}} = \bm{0}$. 
This is consistent with the development in the previous section.

Equations ~\ref{eq:mobility_relation_saddle} and \ref{eq:torque_bal_saddle} form a saddle point system, 
\begin{eqnarray} \label{eq:saddle_point_eqn}
    \begin{bmatrix}
        \mathcal{M}^{\dagger} & \mathcal{B} \\
        \mathcal{B}^{T} & \textbf{R}^{\text{L}\Omega, \text{nf}}
    \end{bmatrix}
    \begin{bmatrix}
        \mathcal{F} \\
        \hat{\bm{\Omega}}
    \end{bmatrix}
    =
    \begin{bmatrix}
        \begin{pmatrix}
            \hat{\textbf{U}} \\
            \bm{0}
        \end{pmatrix} \\
        \skew{-4}\hat{\textbf{L}}^{\text{ext}} - \textbf{R}^{\text{LU}, \text{nf}}\cdot \hat{\textbf{U}}
    \end{bmatrix} ,
\end{eqnarray}
to compute the far-field hydrodynamic forces, torques, and the rotational velocities from the the prescribed translational velocities and instantaneous particle configuration.
Going forward, we let $\skew{-4}\hat{\textbf{L}}^{\text{ext}} = \bm{0}$.
Benzi et al. \cite{Benzi2005-ir} provide an elucidative overview of saddle-point systems and their solution strategies. 
Here, the rotational velocities act as Lagrange multipliers that enforce the torque balance.
In our experiments, the particles experienced external forces arising from a stiff harmonic trap, such that their translational velocities are maintained at prescribed values.
For this reason, we circumvent computing the force balance by asserting that the translational velocities are known.

We time-integrate the linear and rotational particle velocities, $\mathcal{U}$, to get the particle trajectories and orientations.
Throughout this work, Eqn.~\ref{eq:probe_moving_interaction} is a two-body analytical ``theory'' that is obtained by truncating the multipole expansion at the level of hydrodynamic torques.
In comparison, Eqn.~\ref{eq:saddle_point_eqn} is a ``simulation'' method that approximately accounts for many-body hydrodynamic interactions.
\clearpage
\section{\label{sec:results}Results}
\subsection{Fundamental interactions}
\begin{figure}[h!]
\centering
\includegraphics[width=\columnwidth]{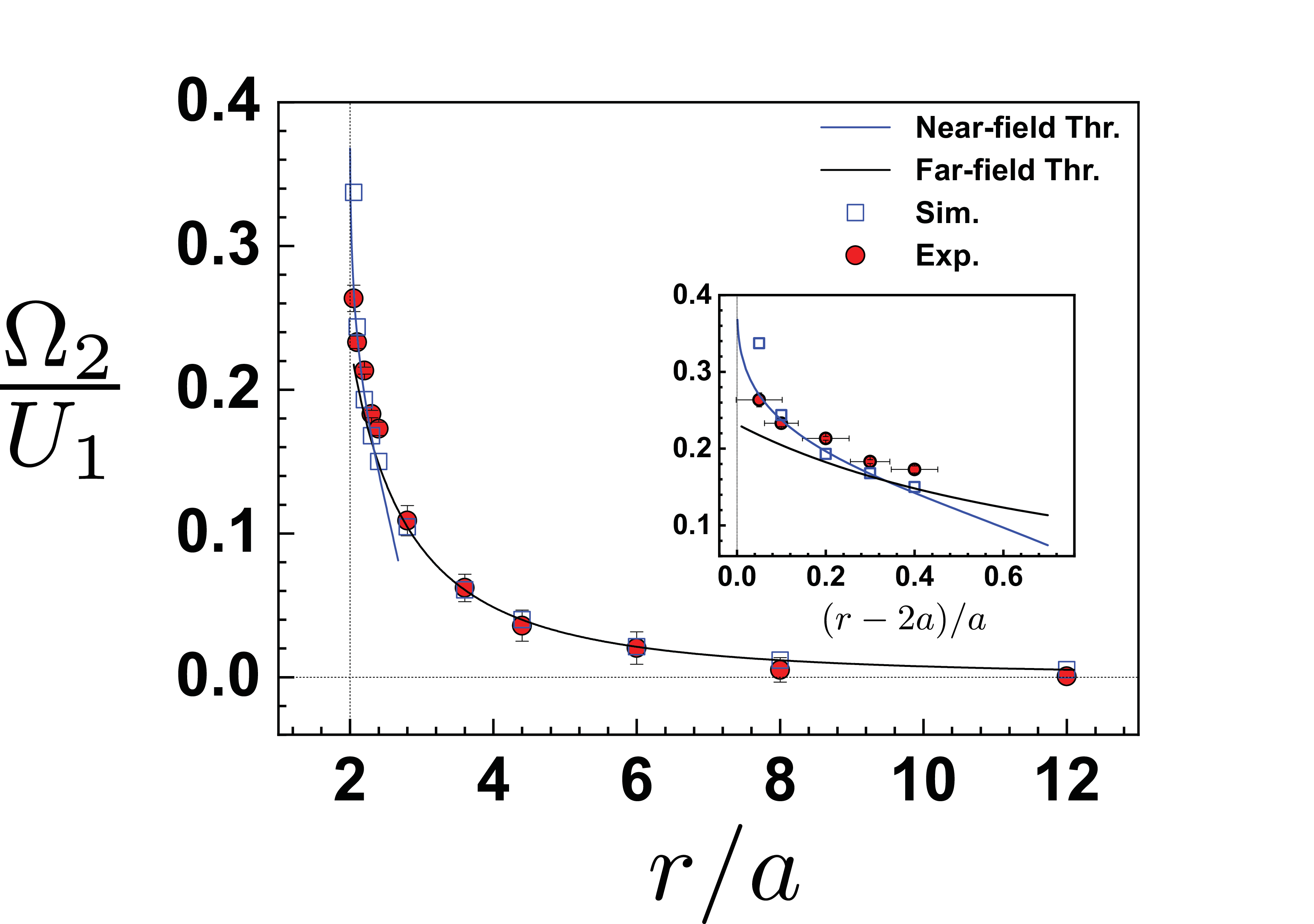}
\caption{\label{fig:fig2} For a pair of spherical particles of radius $a=2.5$ \textmu m, imposing a circular trajectory on one of the spheres (particle 1) generates solvent-mediated forces that rotate a probe (particle 2).
The moving particle translated with speed $U_1 = 35-200$ \textmu m/s.
See Supplemental Movie 2 for a video of the experiment.
We measured the induced rotational velocity of the probe, $\Omega_2$,
as a function of center-to-center distance (``experiment", filled red circles). Positive values indicate counter-clockwise rotation. 
We compared our results to microhydrodynamic theory, Eqn.~\ref{eq:probe_moving_interaction}, that assumes the spheres are widely separated (``far-field theory", solid black line). At small pair-separation distances where the spheres are nearly touching, we use an asymptotic expression, Eqn.~\ref{eq:near_field_interaction} (``near-field theory", solid blue line). 
We also computed the induced rotational velocities using our hydrodynamic simulation method, Eqn.~\ref{eq:saddle_point_eqn} (``simulation", blue open squares).
(Inset) The induced rotational velocities of the probe expressed as a function of the surface-to-surface distance, corresponding to center-to-center distances $r = 2.0a - 2.8a$.
}
\end{figure}
\subsubsection{Two-body interaction}
We measured the dependence of the induced rotational motion of the tracer probe on the pair-separation distance between it and a translating neighbor (Fig.~\ref{fig:fig2}).
Our hemispherical fluorescent labels enabled tracking the probe's orientation to compute its angular displacement, $\theta$ (see Supplemental Movie 1). 
To obtain a reliable measurement of the induced rotational velocity, $\Omega_2 = a\skew{4}{\dot}{\theta}$, at a given pair-separation, $r$, we drove a neighboring particle in a circular orbit for several periods at a range of speeds, $U_1 = 35-200$ \textmu m/s. 
All experiments were conducted at large optical trap stiffness to ignore Brownian fluctuations in the translational motion of particles, focusing solely on their rotational motion. 
The evolution of the angular displacement of the probe particle, measured under varying pair separations and translational velocities, is detailed in Sec.~\ref{sec:appendix_exp_angular displacement}. 
We discovered that fluid-mediated interactions can be detected even at a pair separation of 30 \textmu m, which is approximately 12 particle radii. 
This is significant because the disturbance flow has decayed over 12 particle radii, and the torque fluctuations on the probe become more dominant (see Supplemental movie 2 and 3). 
The rotational polar angle increased linearly with time (see Fig.~\ref{fig:2body_theta}), corresponding to a constant angular velocity.
The time-derivative of the rotational polar angle was determined by performing a minimum of 40 independent measurements, ensuring a reliable measurement of the probe particle's rotational velocity.

At low Reynolds numbers, the rotational velocity is linearly proportional to the translational velocity. 
As such, we collapsed the data by normalizing with respect to the imposed speeds.
In Fig.~\ref{fig:fig2}, we show that the induced rotational velocity of the probe decreases with increasing separation distance (see red circles).
We adopted a microhydrodynamic framework to confirm that solvent-mediated forces give rise to these observations.
We solved the Stokes equations, applying a far-field expansion for ``large'' pair separations, representing each particle in the pair as a point force with corrections for finite-size, namely the hydrodynamic torque and stresslet. 
Accounting for the hydrodynamic forces and torques only, we computed an analytical expression for the induced angular velocity of the freely-rotating probe in the presence of a translating particle with prescribed velocity, Eqn.~\ref{eq:probe_moving_interaction}. 
Only translational motion that is perpendicular to the vector of separation generates rotational motion on the probe.
The moving particle generates a disturbance flow that decays as $\mathcal{O}(r^{-1})$, whose rotational piece is the antisymmetric part of its gradient. 
Thus, to leading order, we expect the interaction to behave as $\mathcal{O}(r^{-2})$, corresponding to the probe's entrainment in the surrounding flow.
Mechanically, the hydrodynamic force, which maintains the prescribed motion of the translating particle, propagates through the fluid and exerts a torque.
Equation \ref{eq:far_field_interaction} describes the strength of this interaction as a function of the separation distance (see solid black line of Fig.~\ref{fig:fig2}). 
We find excellent agreement between the measured rotational velocity and our expression for the far-field interaction for $r > 2.4a$.
The fact that we can detect induced rotational motion at $r$ = 30 \textmu m of separation demonstrates the exquisite precision and sensitivity of our measurements.
To emphasize, for the smallest speed of the moving particle, $U_1 = 35$ \textmu m/s, the hydrodynamic force is roughly $F_1^H \approx 6 \pi \eta a U_1 \approx 2$ pN.
This force then undergoes a $\mathcal{O}(r^{-2})$ reduction as it transmits through the fluid and rotates the probe. 
In the following section, we will find that this sensitivity is crucial for measuring three-body HIs.
At smaller separation distances, our far-field approximation becomes less accurate because we have used the first two multipoles, whereas an infinite number of moments is required to resolve the geometric details of the near-field interaction between the two spheres.
In this limit, we arrive at an asymptotic result for the induced rotational velocity, Eqn.~\ref{eq:near_field_interaction} (see solid blue line of Fig.~\ref{fig:fig2}). 
Again, we find that the experimental data agree very well with microhydrodynamic theory, overall highlighting the extreme precision with which we can measure the translation-rotation hydrodynamic coupling.
In comparison, measuring the translation-translation coupling is more difficult, because the displacements of the particle center-of-mass from the laser position are restricted to a harmonic potential whose width is set by the trap stiffness.
Even with weak trapping, translational displacements weakly report hydrodynamic forces, except in the near-field regime of interaction (see ``SLB only" control data of Fig.~3 in Ref.~\cite{xu2023dynamic}).
There are no such attenuating torques on rotational motion, and thus any observed rotational velocity is a direct signal of fluid-mediated disturbances.

We have only considered the hydrodynamic forces and torques in our far-field expansion, yielding useful expressions, Eqs.~\ref{eq:far_field_interaction} and \ref{eq:near_field_interaction}, for the two-body interaction between the moving particle and rotating probe. 
We also computed the contribution from the particle stresslets using our our simulation method, Eqn.~\ref{eq:saddle_point_eqn}, in Fig.~\ref{fig:fig2} (see blue squares).
They align with the predictions provided by the far-field and near-field theories, showing that the the symmetric force dipoles negligibly impact the predicted rotation rate of the probe, beyond the prediction from the forces and torques.
Later, however, we will find that accounting for the particle stresslets becomes important for resolving multiparticle hydrodynamic interactions.

\begin{figure*}[h!]
\centering
\includegraphics[scale=0.14]{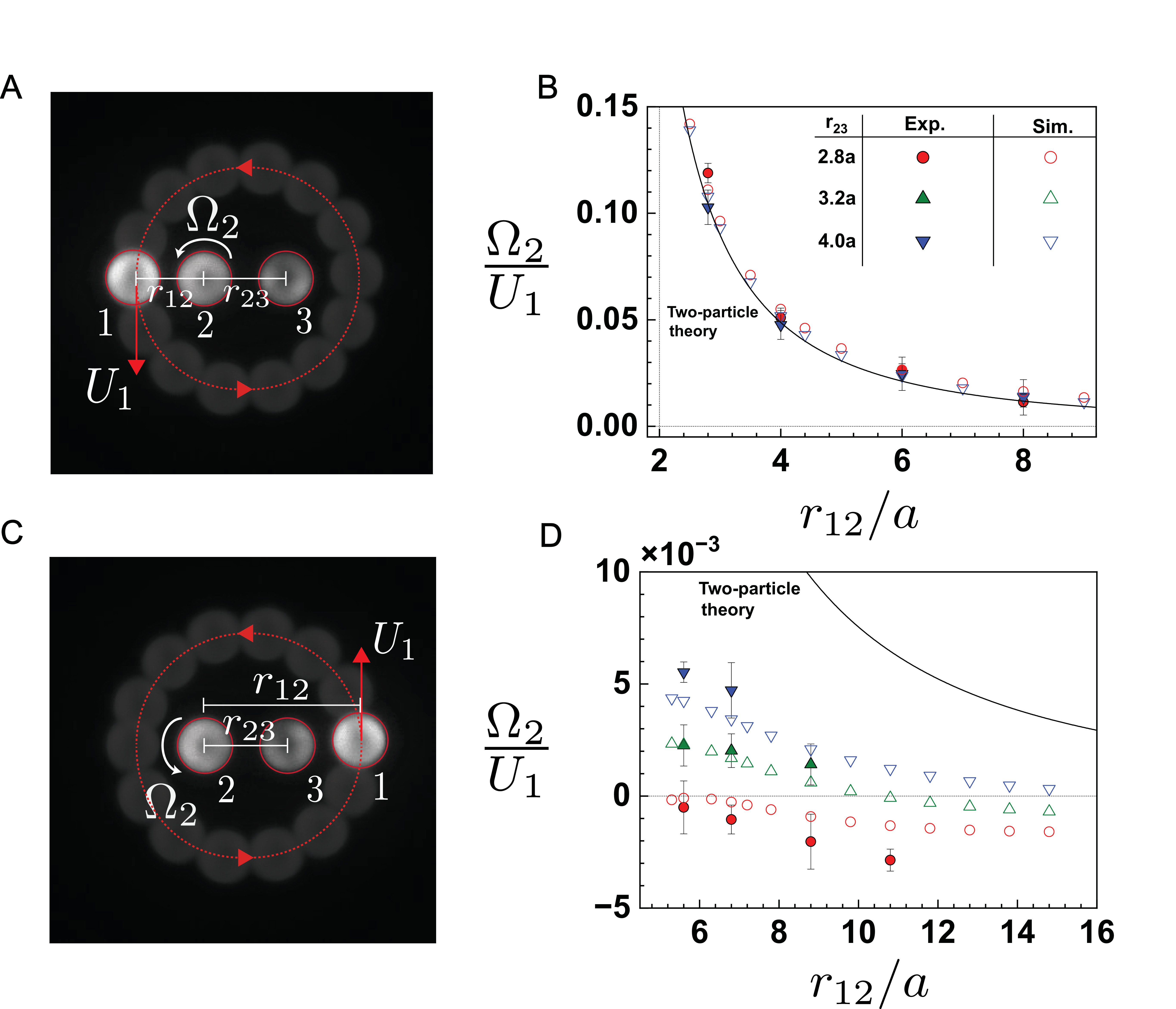}
\caption{\label{fig:fig3} Rotational mobility of a probe exhibits significant variation in three-particle configurations. 
We translated a particle in a circular orbit around two stationary particles.
All radii are $a = 2.5$ \textmu m.
See Supplemental Movie 4 for a video.
(A) A specific three-body configuration of a line of particles, in which the translating particle (labeled ``1'') interacts most directly with the probe (labeled ``2'') in the presence of a third particle (labeled ``3'').
The distance between particles 1 and 2 is $r_{12}$, and between particles 2 and 3 is $r_{23}$. 
We translated particle 1 with speed $U_1 = 100$ and $200$ \textmu m/s, and we measured the rotational velocity of particle 2, $\Omega_2$.
(B) The dependence of $\Omega_2$ on $r_{12}$ at various $r_{23}$.
Positive values indicate counter-clockwise rotation.
Filled symbols are results from experiments, and open symbols from Eqn.~\ref{eq:saddle_point_eqn}.
For clarity, we have only included data at $r_{23} = 2.8a$ and $r_{23} = 4.0a$. 
See Fig.~\ref{fig:3body123_main} for a version of this plot with data at $r_{23} = 3.2a$ and simulation points at $r_{23} = 8.0a$.
For reference, we have also plotted the far-field interaction in the absence of the third particle, Eqn.~\ref{eq:probe_moving_interaction} (solid black line, ``Two-particle theory'').
(C) A specific three-body configuration of a line of particles, where particle 3 screens the interaction between particles 1 and 2. 
(D) The rotational velocity of the probe as a function of $r_{12}$ in the presence of the third particle.
The two-particle theory (solid black line) performs poorly, revealing the significance of the screening posed by particle 3.}
\end{figure*}
\clearpage
\subsubsection{Three-body interaction}
Having now an understanding of the pair hydrodynamic interaction, we introduced multibody effects by adding a third particle and measuring the resulting rotational mobility of a probe (see Fig.~\ref{fig:fig3}A for a sample schematic).
We translated a particle in counter-clockwise, circular orbits (labeled ``1'') about the midpoint of two stationary particles: a probe (labeled ``2'') and a third particle (labeled ``3'').
We fixed the position of the probe with prescribed separation distances from particle 3, $r_{23}$. 

The effect of the third particle was significant (see Supplemental movie 4). 
In the absence of the third particle, the angular displacement of the probe consistently increased in the counter-clockwise direction due the moving particle's orbit, as a function of their separation, $r_{12}$ (Fig.~\ref{fig:2body_3body_comparison}A). 
In contrast, when the third particle was introduced, the probe exhibited periods of stagnation or even brief clockwise reversal (Fig.~\ref{fig:2body_3body_comparison}B).
In general, this effect was observed at the largest values of $r_{12}$ that we tested.
When plotting the time-dependent angular displacements of the probe and the third particle together, their mutual influence becomes distinguishable (see Supplementary Fig.~\ref{fig:2body_3body_comparison}). 
When one particle undergoes rapid rotation, the rotation of the other particle is restricted, exhibiting a symmetric relationship (Fig.~\ref{fig:3body_theta}A). 
Consequently, when the angular velocity of one particle reaches its maximum, the other particle shows the minimum angular velocity (Fig.~\ref{fig:3body_theta}B). 

We focused on analyzing two instantaneous configurations attained by the experiments, one in which the moving particle was nearest to the probe (Fig.~\ref{fig:fig3}A) and another where it was directly intervened by the third particle (Fig.~\ref{fig:fig3}C). 
In the former case, the dependence of the probe's rotational velocity on its separation from the moving particle, $r_{12}$, qualitatively follows our far-field approximation of the two-particle interaction, Eqn.~\ref{eq:far_field_interaction} (compare filled symbols to solid black line in Fig.~\ref{fig:fig3}B). 
These observations are supported by our numerical calculations that account for many-body HIs, which agree with the measurements (compare filled and open symbols in Fig.~\ref{fig:fig3}B).
Both the experimental and theoretical data show that, at increasing $r_{23}$, the pair interaction between the probe and the moving particle approaches Eqn.~\ref{eq:far_field_interaction}, but is consistently larger by a small value.
Therefore, in this configuration, the probe is slightly more mobile in the presence of the third particle, especially at smaller $r_{23}$.
Our measurement is a real three-body effect whose sensitivity is enabled by the torque-free nature of the optical tweezers.
In this specific configuration, the disturbance flow generated by the moving particle resembles that of the purely two-particle interaction. 
However, adding the third particle introduces another obstacle the fluid flow. 
The induced rotation of the probe diverts fluid into the gap between itself and particle 3. 
On average, this ``backflow'' velocity, through the narrow channel of smallest width $r_{23}-2a$, increases with smaller $r_{23}$ to maintain the flow of solvent mass that is driven by the moving particle.
The faster backflow applies greater shear on the probe, resulting in faster spin.
This effect persists for all separations $r_{12}$ and $r_{23}$ we have analyzed (see also green and black points in Fig.~\ref{fig:3body123_main}), demonstrating the long-ranged nature of the disturbance flows.

The rotational velocity of the third particle in Fig.~\ref{fig:fig3}A is equivalent to that of the probe in the configuration shown in Fig.~\ref{fig:fig3}C (labeled ``2''). 
While particle 3 in Fig.~\ref{fig:fig3}C, which is closest to the moving particle, experiences a minor enhancement to its rotational mobility due to the probe (Fig.~\ref{fig:fig3}B), we find that the probe's rotational velocity switches directions depending on its distance from the moving particle (see Supplemental Movie 4). 
Both the experimental and simulation data support this: at a given distance $r_{23}$, there is a distance $r_{12}$ in which counter-clockwise rotation, $\Omega_2 >0$, transitions to clockwise, $\Omega_2 <0$ (compare open and filled symbols in Fig.~\ref{fig:fig3}D).
The interaction between the probe and the moving particle is an induced interaction that is much weaker than the purely two-body interaction (compare solid black line to symbols in Fig.~\ref{fig:fig3}D).
Even though these are $\Omega_2/U_1 \sim \mathcal{O}(10^{-3})$ signals, the agreement between experiment and simulation further confirms the precision of using particle rotation as a reporter of many-body HIs.

Building upon our proposed ``backflow'' mechanism in Fig.~\ref{fig:fig3}B, 
the probe experiences a combination of disturbance flows resulting from the moving particle and the induced rotation on particle 3.
In the gap that separates the probe from particle 3, the flow from the induced rotation on particle 3 points oppositely to the background flow driven by the moving particle.
Therefore, the relative strength of these contributions determines the overall rotation of the probe. 
At a given distance $r_{12}$, the backflow strengthens with decreasing $r_{23}$, as the fluid flows faster through the narrow gap.
There is a critical $r_{23}$ in which the torque from this flow cancels that of the background flow generated by the moving particle, resulting in the observed stagnation of the probe ($\Omega_2 \approx 0$).
Below this critical $r_{23}$, the backflow is sufficiently enhanced to exert greater torque, reversing the probe's spin from counter-clockwise, $\Omega_2>0$, to clockwise rotation, $\Omega_2 <0$ (compare blue, green, and red symbols in Fig.~\ref{fig:fig3}D). 
Put another way, at increasing $r_{12}$, the background flow driven by the moving particle becomes weaker in the vicinity of the probe, and a wider neck between the probe and the third particle (larger $r_{23}$) sufficiently enhances the backflow from particle 3 to switch probe's direction of rotation.
This explains the increasing crossover distance along $r_{12}$, from counter-clockwise to clockwise rotation, for larger $r_{23}$.  
\clearpage
\subsection{Shearing a crystal}
\begin{figure*}[h!]
\centering
\includegraphics[width=\columnwidth]{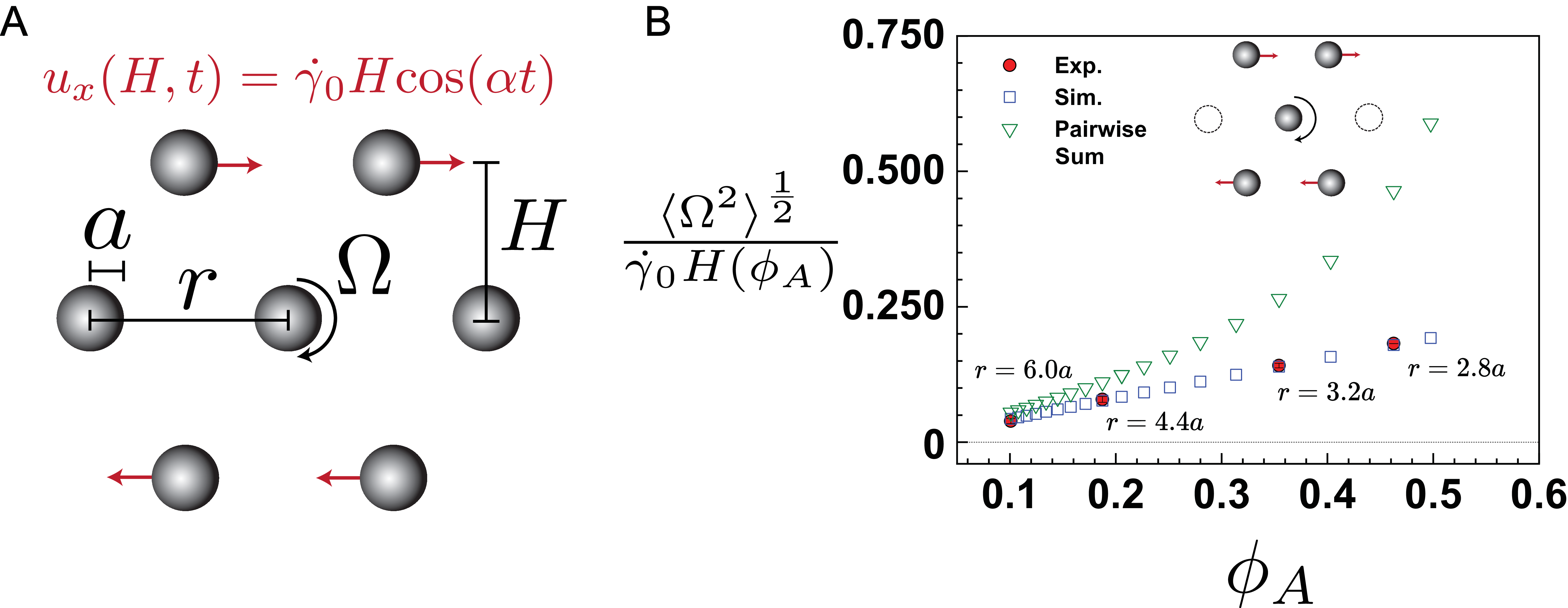}
\caption{\label{fig:fig4} Nearest neighbors hinder rotational mobility during shear of a two-dimensional crystalline array. (A) We trapped seven particles of radius $a = 2.5$ \textmu m into a configuration that mimicked the first coordination layer of a hexagonally-close-packed lattice with pairwise distances $r = 2.8a - 6.0a$. These corresponded to local area fractions $\phi_A = 2\pi a^2(r^{2}\sqrt3)^{-1} \approx 10\% - 46\% $. To model an oscillatory shear strain on the lattice, we imposed a velocity profile in the x-direction of the form $u_x(y,t) = \skew{4}{\dot}{\gamma}_0y\cos(\alpha t)$, with shear-rate amplitude $\skew{4}{\dot}{\gamma}_0 = 0.8\alpha r/H(\phi_A)$ and oscillation frequency $\alpha = 5.5$ Hz. $H(\phi_A) = r(\phi_A)\cos(\pi/6)$ is half of the total height of the lattice. See Supplemental Movie 5 for a video of the experiment.
We measured the time-dependent rotational velocity of the center sphere, $\Omega$. (B) Time-averaged magnitudes of the center sphere's rotational velocity are plotted as a function of the crystalline area fraction.  
Results from experiments (filled red circles) are compared to those obtained from dynamic simulations, using Eqn.~\ref{eq:saddle_point_eqn} (open blue squares).
We also computed rotational velocities from a pairwise sum of the two-body interaction described by Eqn.~\ref{eq:probe_moving_interaction} (open green triangles), where only translating neighbors contribute (see inset graphic). 
The deviation of the pairwise data from our measurements confirms the significance of the bystander particles that scatter fluid disturbances.}
\end{figure*}
\newpage
So far, our focus has been on measuring fundamental, few-body hydrodynamic interactions, where we have successfully measured two- and three-body effects with exquisite precision and accuracy.
Now, we apply these ideas to characterize rotational mobility in a model material with two-dimensional hexagonal packing (Fig.~\ref{fig:fig4}A), whose structure, containing many more particles, gives rise to complex, many-body interactions.

We trapped seven particles in a configuration mimicking the first coordination layer of a hexagonally-close-packed (HCP) lattice. 
All particles were equidistant from one another with spacing, $r$, that was related to the local area fraction, $\phi_A = 2\pi a^2(r^2\sqrt 3)^{-1}$. 
Our aim was to measure the rotational mobility of the center sphere as a function of the packing parameter, $\phi_A$.
We imposed an oscillatory strain on the HCP configuration, where the top and bottom rows of spheres translated in opposite directions along the x-direction according to $u_x(y,t) = \skew{4}{\dot}\gamma_0y\cos(\alpha t)$, and we measured the rotational velocity of the center probe as a function of time (see Supplemental Movie 5).
The shear-rate amplitude was set by the imposed velocity gradient, $\skew{4}{\dot}\gamma_0 = 0.8\alpha r/H(\phi_A)$, where $\alpha = 5.5$ Hz was the oscillation frequency and $H(\phi_A) = r(\phi_A)\cos(\pi/6)$ was half of the total height of the HCP lattice.
We quantified rotational mobility using the time-averaged magnitude of the rotational velocity, $\langle \Omega^2 \rangle^{1/2}(\phi_A) = \left( \frac{1}{T}\int_{t}^{t+T}\Omega^2(t'; \phi_A)\text{d}t' \right)^{1/2}$, defining the period to be $T = 2\pi/\alpha$. 
Finally, in the inertialess limit, the rotational velocity is linearly related to the applied motion on the moving particles. 
To obtain a measurement that was independent of the applied motion, we normalized our time-averaged rotational velocities with respect to the imposed velocity amplitude at a given area fraction, $\skew{4}{\dot}{\gamma}_0H(\phi_A) = 0.8\alpha r(\phi_A)$. 

The results of the time-averages are presented in Fig.~\ref{fig:fig4}B as a function of the HCP lattice area fraction, alongside those from two simulation approaches (see filled red circles and open symbols). 
In one method, we applied a pairwise summation of our two-body analytical expression for the rotation rate, Eqn.~\ref{eq:far_field_interaction}, for each HCP lattice we considered (open green triangles). 
As a result, only translating particles with nonzero velocity were considered.
In the other, we used our hydrodynamic simulation method, Eqn.~\ref{eq:saddle_point_eqn}, including the particle stresslets in our multipole expansion (open blue squares).
We observe excellent agreement between the experiments and simulations using Eqn.~\ref{eq:saddle_point_eqn} (compare open blue squares to filled red circles), whereas the predictions from the pairwise summation exceed the measurements, even at low area fractions (compare open green triangles to filled red circles).
These results highlight the significance of the ``bystander'' particles on the scattering of multiparticle HIs, and the errors incurred in a theory of simple pairwise summation of two-body interactions.
As the top and bottom rows of spheres in the HCP lattice undergo relative translation, they produce a disturbance field in the solvent that resembles planar Couette flow. 
At each instant during the oscillatory shear, the translating particles have velocities with parallel and perpendicular components to the line-of-centers connecting them to the stationary particles. 
Induced rotations arise from motion along the perpendicular components, which our pair summation method solely considers.
However, relative motion along the parallel components simultaneously push and pull solvent out of or into the interfacial gaps separating them from their neighbors, setting up an extensional flow. 
As rigid particles cannot deform, the fluid responds by exerting symmetric-force dipoles, Eqn.~\ref{eq:stresslet_constraint}, that hinder the particles' mobility.
This effect is enhanced at larger area fractions due to the closer packing (see labeled HCP pair distances in Fig.~\ref{fig:fig4}B), requiring more work to pump solvent through the smaller open-pore space.
In comparison, pairwise summation of Eqn.~\ref{eq:probe_moving_interaction} leads to divergent behavior at large $\phi_A$ because the closer packing entails a stronger induced rotation, while neglecting the attenuating effect from the induced stresslets.
Our results highlight the role of multi-body hydrodynamics and offers an effective approach for characterizing regularly packed crystal structures.

\subsection{Inverse problem: Specifying the angular velocity under a constrained trajectory}
\begin{figure*}[h!]
\centering
\includegraphics[width=\columnwidth]{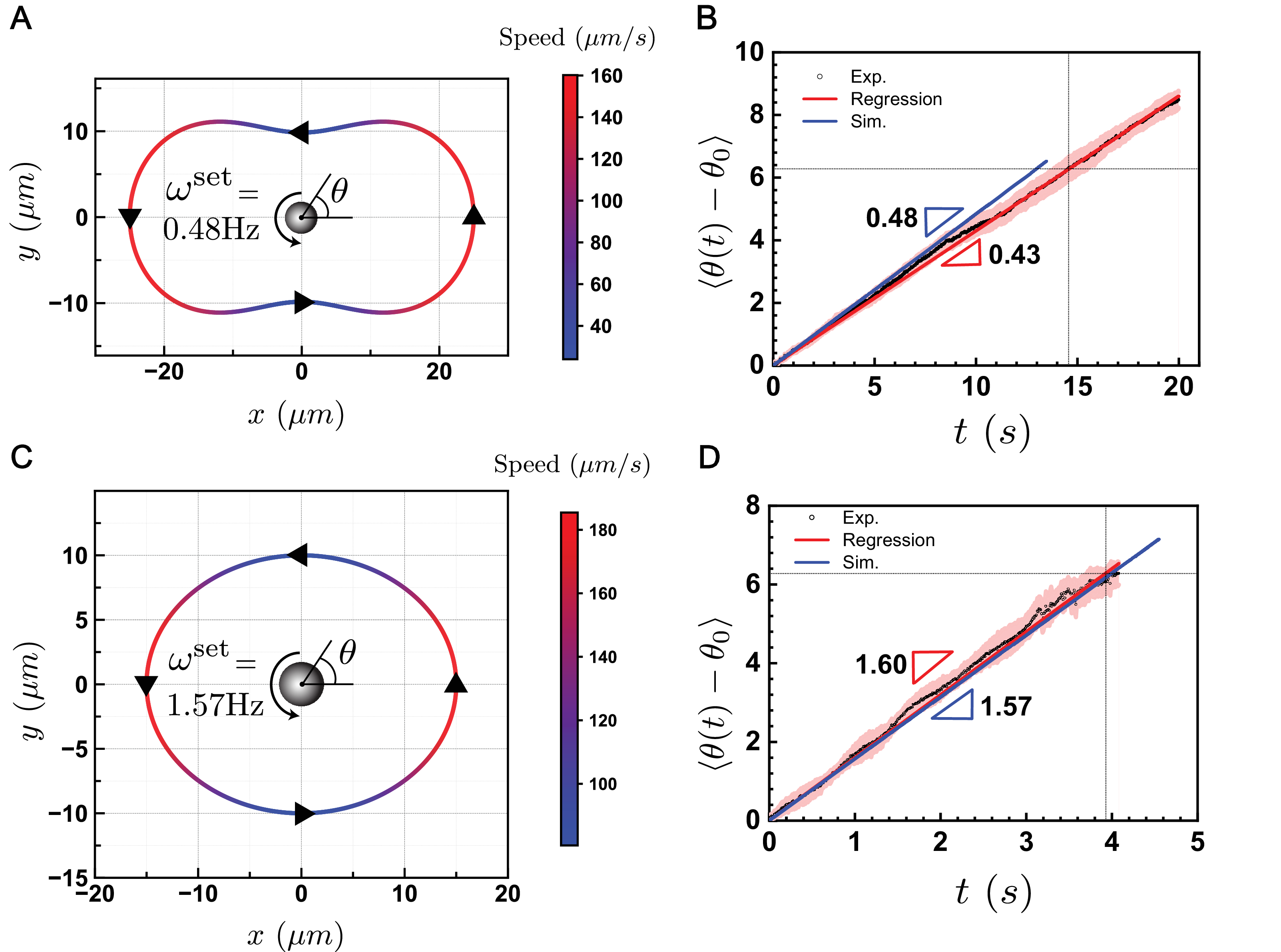}
\caption{\label{fig:fig5} From a desired angular velocity of a probe ($\omega^{\text{set}})$, we computed the trajectory of a nearby moving particle that follows non-circular paths.
Here, we considered (A) a Cassini oval and (C) an ellipse. 
The instantaneous translational velocity of the moving particle is a function of the trajectory's contour (see Sec.~\ref{sec:appendix_inverse_formulation} for computational details), where blue and red colors indicate slower and faster speeds, respectively. Arrows indicate the direction of the moving particle. Using these computed trajectories, we performed laser tweezer experiments to verify our calculations (panels B and D). 
See Supplemental Movie 6 for a video of the experiments.
We measured the time-dependent polar angle, $\theta(t)$, of the probe in radians (black points), applying a moving average and regressing to obtain the angular velocity (solid red lines). The intersection of the gray lines indicates when the probe completes one period of rotation in the experiments. 
We also performed dynamic simulations by solving Eqn.~\ref{eq:saddle_point_eqn} with the computed trajectories (solid blue lines).}
\end{figure*}
We move away from the ``bottom-up'' approach we have presented, using particle rotations as a reporter for fluid-mediated disturbances.
The robustness of our experimental methodology enables us to use our theoretical framework to manipulate HIs to our desired effect.
Instead of specifying a translational velocity on the moving particle and observing the induced rotation rate of the probe, we now wish to do the inverse: specify the probe's angular velocity and determine the required translational velocity of the moving particle. 
We aim to demonstrate the potential of our theoretical framework to direct particle orientation, achieved through optical manipulation of local fluid flows.

We return to the two-particle setup, as shown in Fig.~\ref{fig:fig2}A. 
Motion that is perpendicular to the vector of separation between the two particles determines the angular velocity, leaving a degree of freedom along the parallel component of motion that allows for any number of eligible trajectories of the moving particle.
A circular trajectory is the simplest of these. 
Along the circumference, the particle moves perpendicularly to its separation from the probe with constant speed, because of the fixed radius of separation. 
However, for an arbitrary non-circular trajectory, the required speed of the moving particle is dependent on its current position. 
Larger separations from the probe require faster speeds to compensate for the spatial decay of the hydrodynamic interaction. 

As case studies, we imposed that the moving particle followed two non-circular trajectories, a Cassini oval and an ellipse, determining its velocity along the contours to give a desired angular velocity of the probe.
Having the particle move in non-circular trajectories, while inducing the desired rotation rate of the probe, necessitates additional shape-related constraints.
In Sec.~\ref{sec:appendix_inverse_formulation}, we derive one such ``inverse formulation'', arrived from the original saddle-point system in Eqn.~\ref{eq:saddle_point_eqn}, and employ a two-step time-integration scheme to constrain the moving particle onto the desired trajectory shape.
Panels A and C of Fig.~\ref{fig:fig5} present the results of our theoretical calculations for the velocity of the moving sphere along a Cassini oval and an ellipse, where we have specified the angular velocity of the probe to be 0.48 Hz and 1.57 Hz, respectively. 
These values were chosen to ensure that the predicted speeds were within the saturation limits of the optical tweezers, while suppressing thermal fluctuations.
For both trajectories, our calculations show that, indeed, the particle moves faster when it is farther away from the probe to maintain the desired angular velocity (compare red to blue colors along the contours). 
To verify our results, we performed laser tweezer experiments using the trajectories from panels A and C of Fig.~\ref{fig:fig5} (see Supplemental Movie 6).
Panels B and D are measurements of the induced angular displacement of the probe during the experiment (black points), whose slope is a measure of the angular velocity (red lines). 
We corroborated our results, using dynamic simulations that solve Eqn.~\ref{eq:saddle_point_eqn} with the trajectories predicted from the inverse formulation (blue points). 
We observed excellent agreement between the angular velocities set by the simulation and those obtained from experiment.
We have demonstrated that microhydrodynamic theory not only predicts ``bottom-up'' measurements of fluid-mediated forces, but one can use it to manipulate the governing fluid flows for a targeted outcome.

\section{\label{sec:conclusion}Conclusion}
In this study, we developed an innovative experimental method to quantify precisely fluid-mediated multi-body interactions by inducing translation-rotation hydrodynamic coupling between trapped colloids.
Optical tweezers do not apply external torques, enabling extremely sensitive measurements of angular displacements that report fluid-mediated interactions.
For example, we measured the induced rotational velocity of a stationary tracer probe due to a nearby, translating particle, and detected fluid disturbances generated by $\sim$2 pN of hydrodynamic force from over 10 particle radii of separation. 
This level of sensitivity was crucial for measuring three-body interactions.
For certain three-body configurations, the rotation of a probe switched directions, depending on the competing influence of induced flows generated by a moving particle and its nearest neighbor.
We further investigated the role of multiparticle HIs by imposing oscillatory strain on a regular, hexagonal lattice of colloidal particles.
Our results revealed the importance of induced stresslet interactions, especially at close packing, where the fluid in the open-pore space exerts resistance to particle rigidity. 

Beyond direct measurements of fundamental hydrodynamic interactions, we used optical tweezers to manipulate the local flows about a tracer probe to control its rotation.
To do so, we derived an ``inverse'' formulation to design viscous fluid flows for a targeted outcome. 
By dynamically controlling particle trajectories using laser tweezers, we successfully tailored pair HIs to induce a desired angular velocity of a probe particle, demonstrating the feasibility of manipulating colloidal-scale flows.
For example, shearing a solution of worm-like micelles forces the polymeric chains to align with the outgoing flow, ultimately lowering the effective viscosity at large shear-rates \cite{Shikata1988,Cates1990-ew}.
From an engineering standpoint, shear-thinning is essential to the delivery of consumer products and polymer-based injection and extrusion-molding processes.
For spherical colloids, which we are concerned with, patchy surfaces introduce tunable, anisotropic pair interactions that are attractive for multifunctional material design and biomimetic reconstitution \cite{van-Blaaderen2006-nb, Glotzer2007-tj, Zhang2004-rw, Kamp2020-xc, Jones2012-tk, Bianchi2011-xp}.
By adjusting patch valency, one can make chains or sheets that are traditionally achieved by elongated or rod-like constituents \cite{Chapman1967-wa, Powers1975-de, Preisler2013-gc, Noya2014-rh, Ferrari2017-uu, Zhao2018-zl}.
Non-equilibrium routes toward patchy colloidal self-assembly enable maneuvering the landscape of possible suprastructures that may be inadmissible from thermal forces alone \cite{Nag2024-ep, Arango-Restrepo2019-ge}.

An extension to our work is to drive a particle to control the position and orientation of a tracer probe under weak trapping. 
Now, the probe exhibits non-negligible fluctuations that present a realistic challenge in our aim of conducting microscopic materials engineering.
Using multiple agents to assemble many colloidal building blocks into a complex structure  requires knowledge of many-body HIs.
Designing the agents' trajectories to handle disturbances and multiparticle interactions in a physically-informed manner is achievable by integrating the Stokesian Dynamics method \cite{Fiore2019-or} in a control-loop algorithm.
Taken together, the ability to measure, predict, and control hydrodynamic interactions may become critical for multifunctional material design by steering the polarity of patchy colloidal surfaces.

This work not only reinforces the strength of microhydrodynamic theory for elucidating hydrodynamic interactions, but also underscores its potential for active manipulation of interactions to achieve desired system behaviors. 
Another area of interest is controlling living materials consisting of motile microorganisms \cite{Yang2025-fn, Massana-Cid2022-ey, Frangipane2018-mh, Arlt2018-qs, Arlt2019-yh}.
Promising recent results of feedback and control of freely-draining active Brownian particles \cite{Quah2024-qj, quah2025intabp} with a physics-informed model strongly suggest that a similar approach can be applied in a fluid mechanical framework.
By integrating microhydrodynamic models with real-time experimental feedback, colloidal agents can direct complex, microbe-generated fluid flows and traverse a rich, non-equilibrium phase space of colony-scale structures \cite{baskaran2009statistical, marchetti2013hydrodynamics, shankar2019hydrodynamics, aditi2002hydrodynamic}.
\section{Acknowledgments}
The authors thank Titus Quah for valuable discussions regarding inverse problem-solving strategies.
S.C.T. is supported by the Packard Fellowship in Science and Engineering.

\section{Appendix}
\subsection{Measurement of rotational diffusivity}
\label{sec:appendix_exp_rotational diffusivity}
Rotational diffusivity, $D_r$ , quantifies how quickly a particle loses its orientational memory due to thermal fluctuations. 
For spherical particles, this transport coefficient is extracted by analyzing the time-dependent autocorrelation function of the orientation vector.
Below, we describe the principles and methodology in detail. 

The orientation vector, $\bm{q}$, was determined by analyzing the brightness profile along the circumference of a spherical particle that was half-coated with a fluorescent marker. 
The Fourier transform of the brightness profile was used to identify the point corresponding to the maximum or minimum intensity. 
The autocorrelation function, $C(t)$, for the observable orientation vector in this setup follows
\begin{eqnarray}
    C(t) = \langle \bm{q}(t) \cdot \bm{q}(0) \rangle = \exp(-D_r t) .
\end{eqnarray}
Supplemental movie 1 shows the time-dependent rotational Brownian motion of a hemispherical fluorescent PMMA particle under very weak optical trap stiffness.
The video clearly demonstrates the precise tracking of both the particle's center and the brightest region surrounding it. 
The autocorrelation of the orientation vector as a function of time can be determined from the experimental observations (Fig. \ref{fig:appendix_a1}). 
The experimentally measured rotational diffusivity was 0.010 rad/s, which agrees with the 2D Stokes-Einstein-Sutherland rotational diffusivity. 
In our experiments, the particles were fluorescently labeled on one side, which likely contributed to a deviation from perfect sphericity. This asymmetry in particle structure can introduce anisotropic interactions with the surrounding fluid, potentially biasing the alignment of particles along the x- and y-directions. 
As a result, the observed rotational dynamics are primarily confined to in-plane rotations, with out-of-plane rotation rarely occurring.

To assess the significance out-of-plane rotations, we trapped a tracer probe and measured the time-dependence of the area of a triangle formed by the brightest points along three concentric circles (Fig.~\ref{fig:triangle}).
We considered two cases, one in which the probe exhibited purely diffusive motion (Fig.~\ref{fig:triangle}A) due to thermal fluctuations, and another where we induced rotational motion by translating a nearby particle in a circular orbit (Fig.~\ref{fig:triangle}B).  
In panels A and B of Fig. \ref{fig:triangle}, we drew three concentric circles at equal intervals from the center of the probe particle (yellow line: particle perimeter, yellow cross: circle center). 
We tracked the brightest point along each circle, comprising the vertex of a triangle whose area we monitored over time (Fig.~\ref{fig:triangle}C).
In purely two-dimensional rotational motion, the area of this triangle would remain constant.
However, out-of-plane Brownian torques produce fluctuations in the area of the triangle (see red points in Fig.~\ref{fig:triangle}C).
In the case of purely diffusive rotational motion, we observe that the area of the triangle fluctuates about a mean value, suggesting that out-of-plane rotational displacements are confined to a potential that possibly results from vertical pinning.
In contrast, when the probe undergoes rotational motion due to hydrodynamic interaction with a nearby translating particle, the fluctuations in the triangular area become smaller, indicating that the fluid flow suppresses out-of-plane rotation (see blue points in Fig.~\ref{fig:triangle}C). 
Particle roughness and slight asymmetry can also impact these results, by influencing the hydrodynamic forces and torques acting on the probe. 
Such factors would constrain rotations to in-plane motion and limit out-of-plane fluctuations.
In the experiments, we rarely observed out-of-plane rotational motion.
Furthermore, the high P\'eclet number along the xy-plane in our experiments ensured that rotational motion in other planes could be safely neglected, as thermal fluctuations were overwhelmed by the fluid flows generated by imposed translational motion of the neighboring particle. 
The strong agreement between our experimental findings and theoretical predictions highlights the precision of our measurement approach and the robustness of our experimental setup in accurately capturing particle rotational dynamics.

\renewcommand{\thefigure}{A\arabic{figure}}
\setcounter{figure}{0}

\begin{figure}[h!]
\centering
\includegraphics[width=15cm]{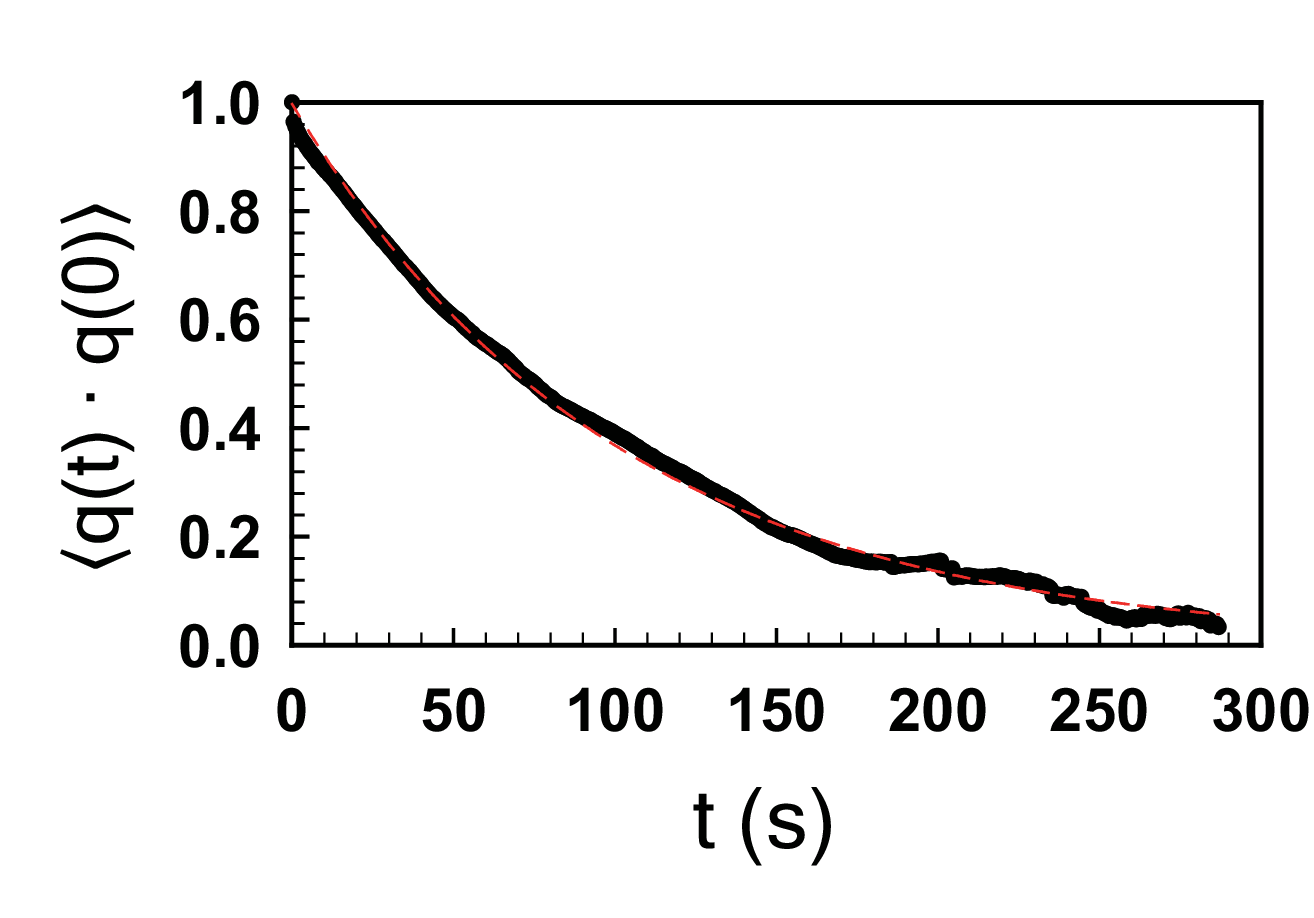}
\label{fig:appendix_a1}
\caption{\label{fig:appendix_a1} The time-dependent autocorrelation function of the orientation vector, $C(t) = \langle \bm{q}(t) \cdot \bm{q}(0) \rangle$, exhibits an exponential decay, which is used to determine the rotational diffusivity, $D_r$. The black solid line represents the experimental data, while the red dashed line corresponds to the regression fit. The theoretical rotational diffusivity was calculated to be $0.0102 \, \mathrm{rad/s}$, using the viscosity of water at room temperature.}
\end{figure}

\begin{figure}[h!]
\centering
\includegraphics[width=15cm]{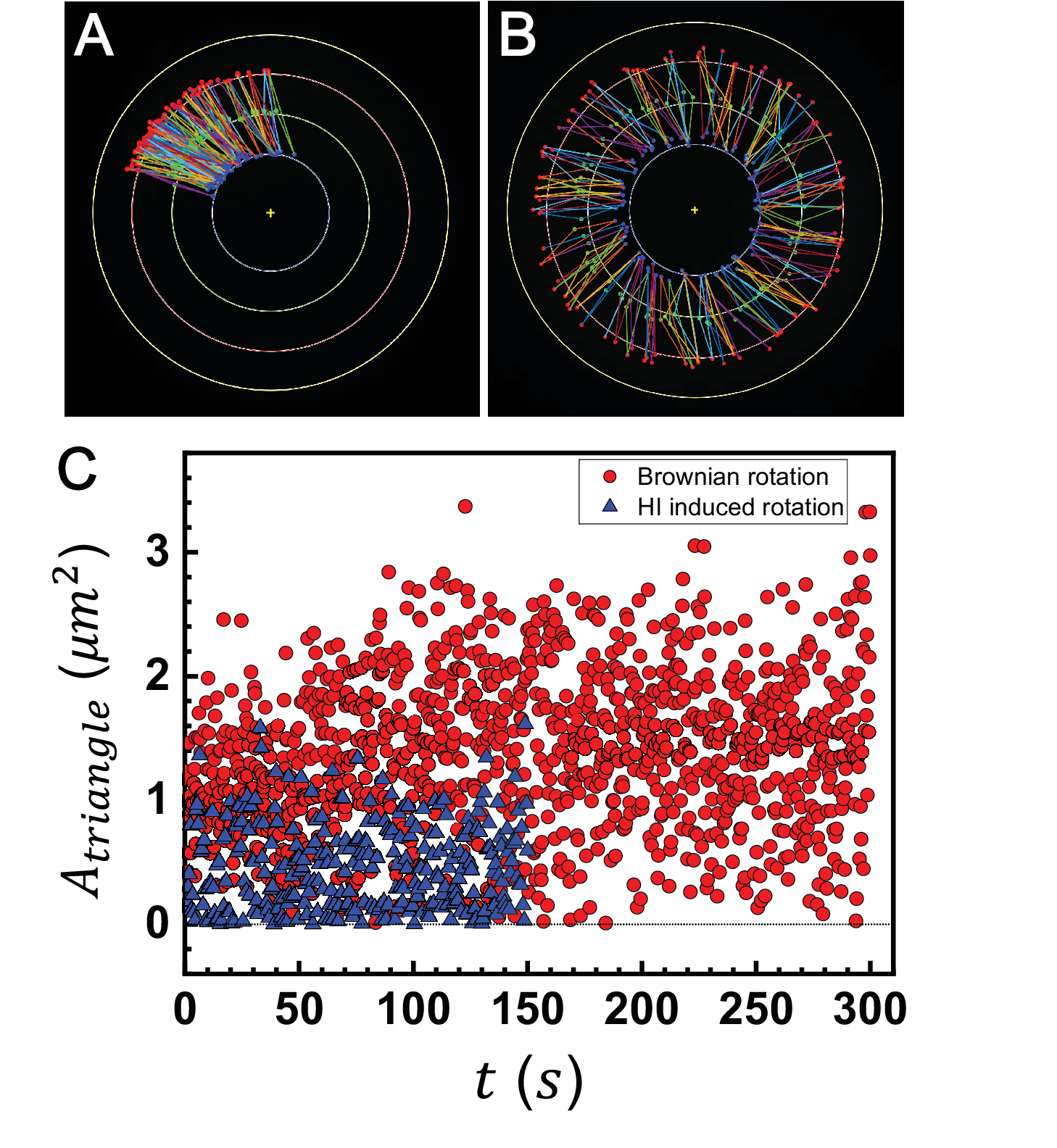}
\label{fig:triangle}
\caption{\label{fig:triangle} Schematic of the method used to analyze out-of-plane rotation under (A) Brownian rotation and (B) induced rotation due to HI with a nearby translating particle. Three concentric circles were drawn at equal intervals from the center of the probe particle (yellow line: particle perimeter, yellow cross: circle center), and the brightest points along each circle were identified and marked. We computed the time-dependent area of the triangle formed by these three points to assess variations due to out-of-plane rotations. (C) Time-dependent fluctuations in the triangular area under Brownian motion (red points) and rotational motion due to HI (blue points).}
\end{figure}

\clearpage
\subsection{Near-field interaction between a rotating probe and a translating sphere of equal size} \label{sec:appendix_nf_interaction}
Analogously to the mobility relation presented in Eqn.~\ref{eq:mobility_relation} of the main text, we use a resistance formulation to relate the hydrodynamic forces and torques to the particle velocities. 
\begin{eqnarray}\label{eq:appendix_res_relation}
\begin{pmatrix}
\hat{\textbf{F}}^{\text{H}}\\
\skew{-4}\hat{\textbf{L}}^{\text{H}}
\end{pmatrix}
=
\begin{pmatrix}
\textbf{R}^{\text{FU}} & \textbf{R}^{\text{F}\Omega}\\
\textbf{R}^{\text{LU}} & \textbf{R}^{\text{L}\Omega}
\end{pmatrix}
\begin{pmatrix}
\hat{\textbf{U}}\\
\hat{\bm{\Omega}}
\end{pmatrix} . 
\end{eqnarray}
For freely-rotating particles, $\skew{-4}\hat{\textbf{L}}^{\text{H}} = \bm{0}$.
Using tabulated, asymptotic expressions for the pairwise resistance tensors \cite{Kim1991-cs}, we solve for the rotational velocities in Eqn.~\ref{eq:appendix_res_relation}, 
\begin{eqnarray}
    \hat{\bm{\Omega}} = -\left( \mathbf{R}^{\text{L}\Omega} \right)^{-1} \cdot \mathbf{R}^{\text{LU}} \cdot \hat{\textbf{U}}
\end{eqnarray}
The rotational velocity of the probe is $\bm{\Omega}_2 = Y^*(\xi)\hat{\bm{r}}_{21} \times \bm{U}_1$, where $\xi = (r_{21} - 2a)/a$.
Reproducing Eqn.~\ref{eq:near_field_interaction} of the main text, the scalar function has the form of
\begin{eqnarray*}
    Y^{*}(\xi) = \prod_{m = 0}^{1}\left[ \sum\limits_{n = 0}^{3}h_{mn}(\xi)\ln(\xi)^n \right]^{2m-1} .
\end{eqnarray*} 
The coefficients, $h_{mn}(\xi)$, are power series in $\xi$ of order $n$,
\begin{eqnarray}
h_{00}(\xi) &= 92.33, \\
h_{01}(\xi) &= 79.4175 + 75.4425 \xi, \\
h_{02}(\xi) &= 22.1436 + 40.4133 \xi + 17.3124 \xi^2, \\
h_{03}(\xi) &= 1.99787 + 5.22378 \xi + 4.20886 \xi^2 + \xi^3, \\
\nonumber \\
-h_{10}(\xi) & = -0.724834, \\
-h_{11}(\xi) &= 17.9685 + 11.0943 \xi, \\
-h_{12}(\xi) &= 8.68238 + 11.327 \xi + 3.76847 \xi^2, \\
-h_{13}(\xi) &= 0.998939 + 1.86469 \xi + 1.08325 \xi^2 + 0.200321 \xi^3.
\end{eqnarray}

\clearpage
\subsection{Results for three-body interaction} 
\subsubsection{Two-body configuration}\label{sec:appendix_exp_angular displacement}
In a two-body configuration, the time-dependent changes in the rotational angle of the probing particle were observed by varying the pair separation and the imposed velocity of the moving particle. In all cases, the rotational angle of the probing particle exhibited a linear relationship with time, and the rotational velocity was calculated using the slope. The probe particle showed faster rotational velocities as the pair separation decreased and as the velocity of the moving particle increased (Fig. \ref{fig:2body_theta}). Experimental videos for each condition can be found in Supplemental Movie 2 and 3.

\begin{figure}[h!]
\centering
\includegraphics[width=15cm]{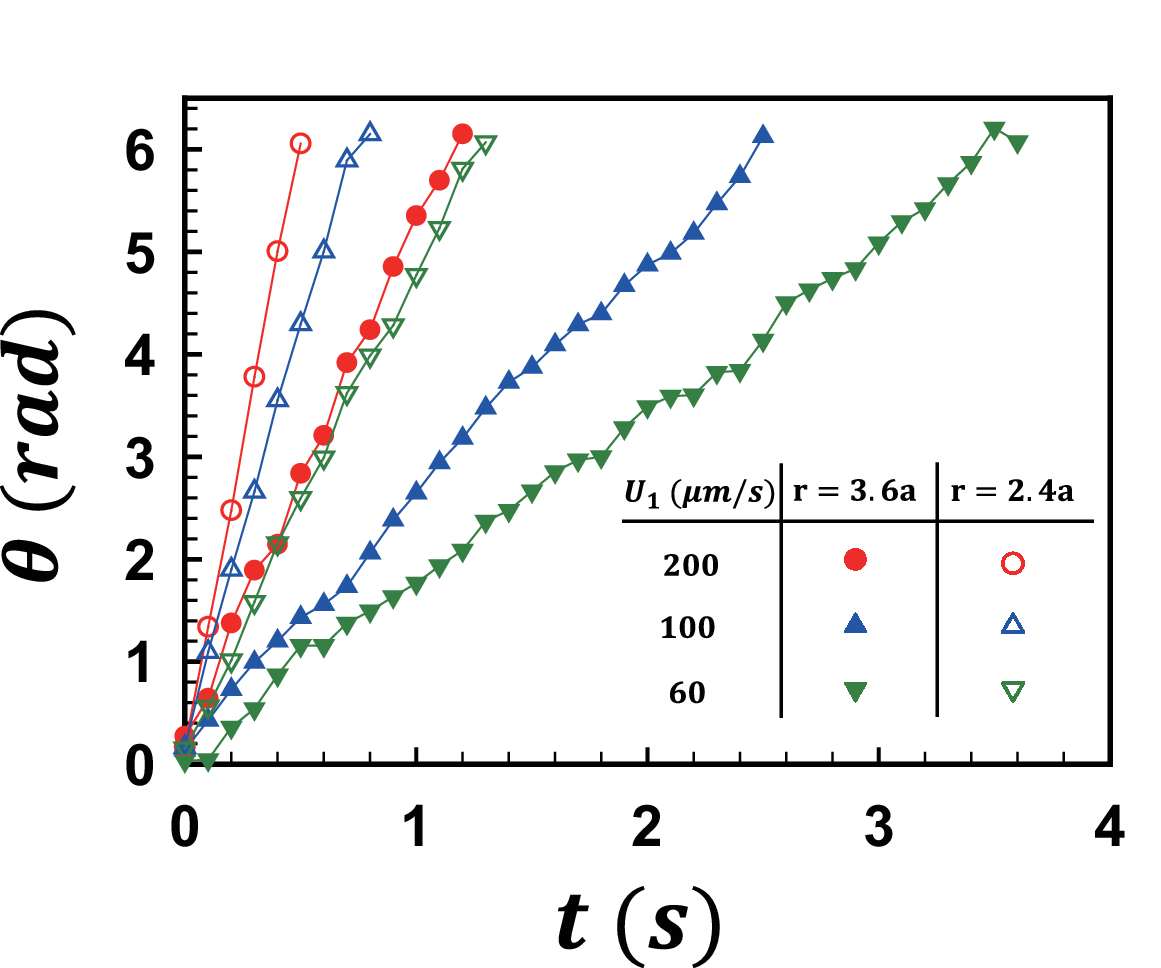}
\label{fig:2body_theta}
\caption{\label{fig:2body_theta} Time-dependent change in the rotational angle of the probing particle at various pair separations and imposed velocities.}
\end{figure}

\subsubsection{Three-body configuration}\label{sec:appendix_exp_angular displacement_three}
The impact of the third particle was evident. Without particle 3, the rotational angle of particle 2 consistently increased in a counter-clockwise direction, following the motion of particle 1 and varying with the distance between particle 1 and particle 2. In contrast, the introduction of particle 3 led to intervals where particle 2’s rotation stagnated or briefly reversed in a clockwise direction (Fig.~\ref{fig:2body_3body_comparison}). In Fig.~\ref{fig:2body_3body_comparison}, the experimental setup fixed the distance between particle 2 and particle 3 ($r_{23}$) at 6 \textmu m, with the minimum distance between particle 1 and particle 2 set at 7 \textmu m. The effect of particle 3 was most pronounced when the distance between particle 1 and particle 2 was greatest, resulting in a reversal of particle 2's rotation. When the rotational angle changes of particle 2 and particle 3 are examined together, their mutual interaction becomes apparent. Specifically, when one particle undergoes rapid rotation, the rotation of the other is suppressed, demonstrating a symmetric dynamic. As a result, the angular velocity of one particle peaks when the angular velocity of the other particle is at its lowest (Fig.~\ref{fig:3body_theta}). Experimental videos for three-body configuration can be found in Supplemental Movie 4.

\begin{figure}[h!]
\centering
\includegraphics[width=\columnwidth]{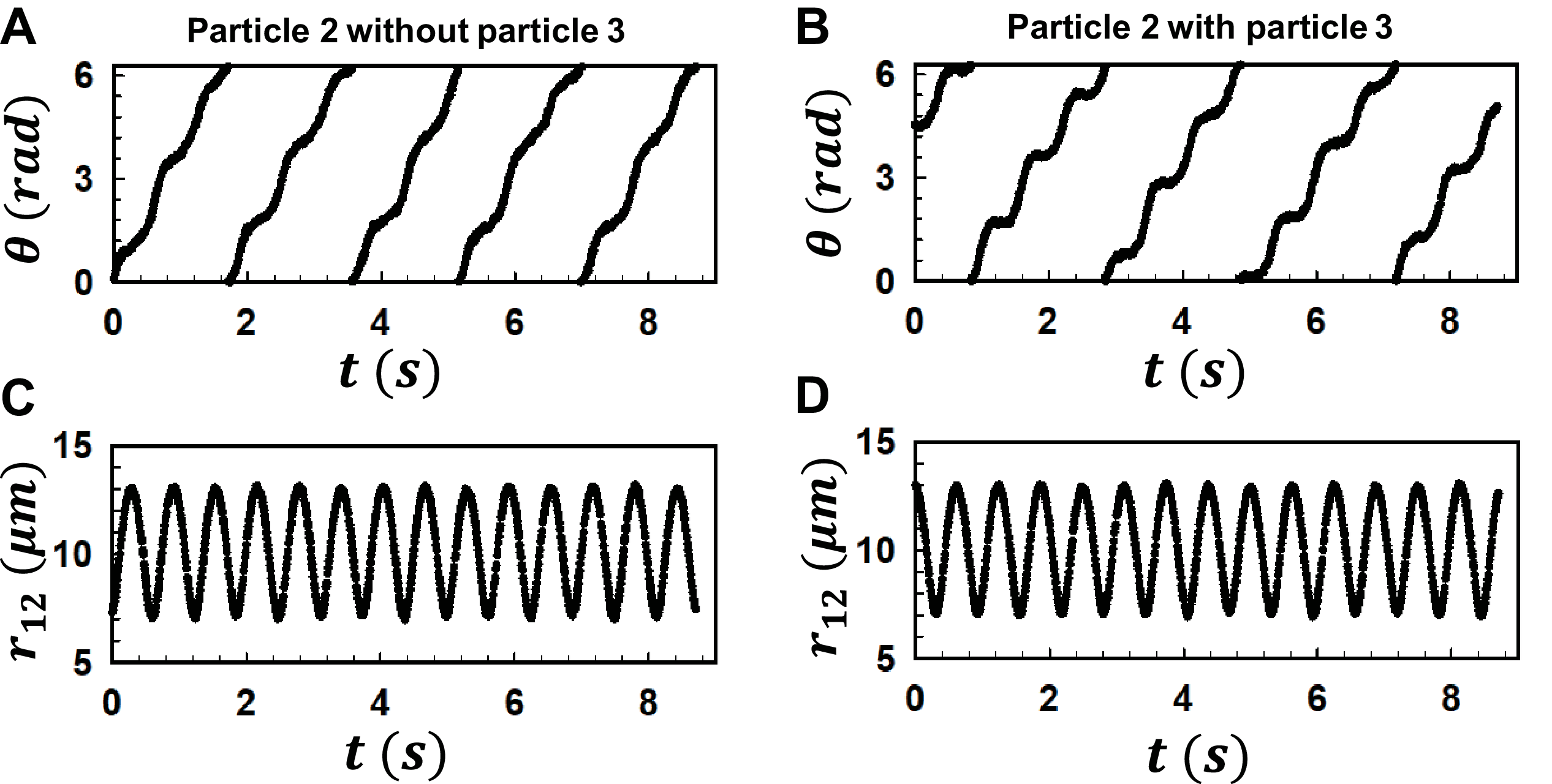}
\label{fig:2body_3body_comparison}
\caption{\label{fig:2body_3body_comparison} Changes in the rotational motion of particle 2 (probing particle) influenced by the presence or absence of a third particle (particle 3) in relation to particle 1 (moving particle).  (A) Time-dependent change in the rotational angle of particle 2 in the absence of particle 3. (B) Time-dependent change in the rotational angle of particle 2 in the presence of particle 3. (C) and (D) Corresponding time-dependent variations in the distance between particle 1 and particle 2 for panels A and B, respectively. Velocity of the moving particle was fixed to 100 \textmu m/s and the distance between particle 2 and particle 3 maintained at 6 \textmu m. The minimum distance between particle 1 and particle 2 was set at 7 \textmu m.} 
\end{figure}

\begin{figure}[h!]
\centering
\includegraphics[width=12cm]{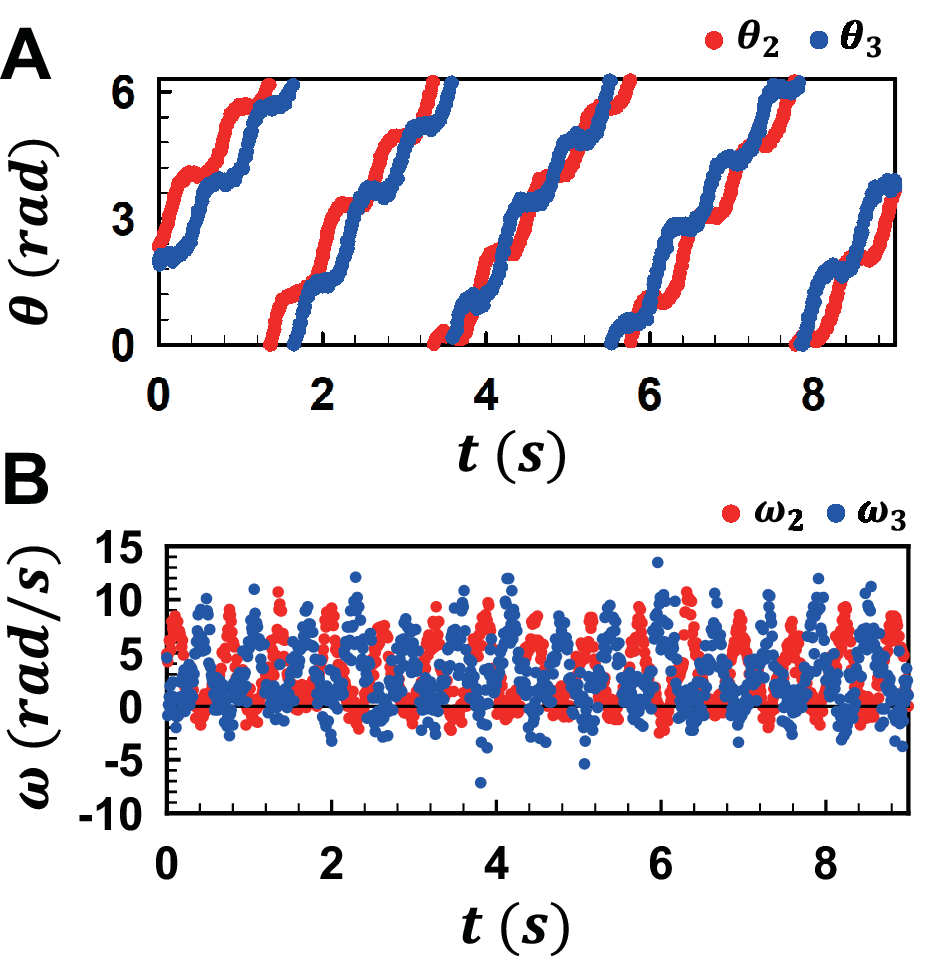}
\label{fig:3body_theta}
\caption{\label{fig:3body_theta} Time-dependent changes in the three-body configuration for particle 2 and particle 3: (A) rotational angle, and (B) angular velocity. Velocity of the moving particle set to 100 \textmu m/s and the distance between particle 2 and particle 3 maintained at 6 \textmu m. The minimum distance between particle 1 and particle 2 was set at 7 \textmu m.
}
\end{figure}

\label{sec:3body_123_main_appendix}
\begin{figure}[h!]
\centering
\includegraphics[scale=0.5]{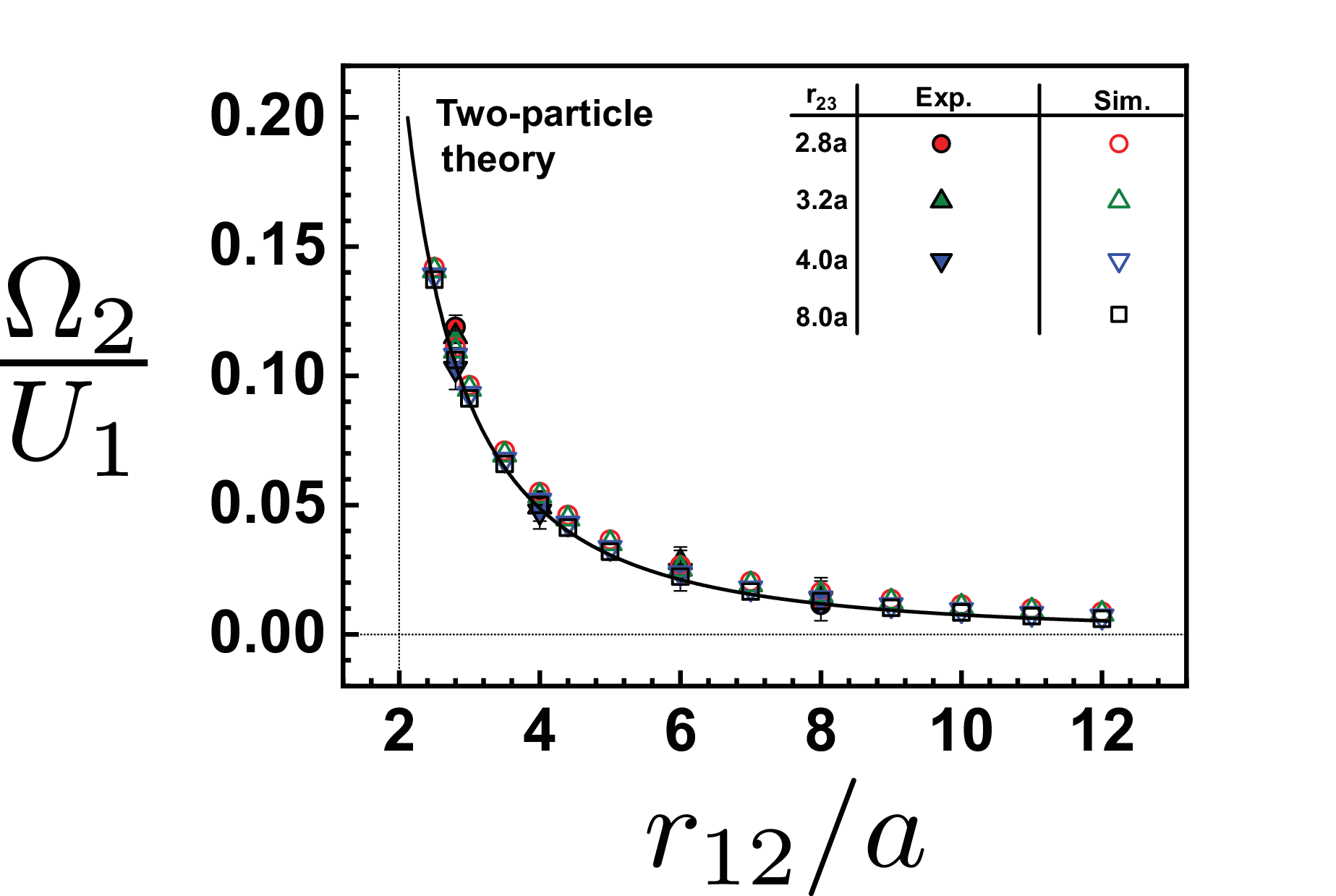}
\caption{\label{fig:3body123_main}
Expanded version of Fig.~\ref{fig:fig3}B with data from experiments and simulations at $r_{23} = 3.2a$ (open and filled green triangles).
We have also included calculations from simulations at $r_{23} = 8.0a$ (open black squares).}
\end{figure}

\clearpage

\subsection{Inverse formulation for two spheres} \label{sec:appendix_inverse_formulation}
We consider the problem of specifying the rotational velocity of a particle and finding the required translational velocity of a neighboring particle. 
To close the problem, we need to impose additional constraints on the translating particle's trajectory.

Starting from the saddle point problem, Eqn.~\eqref{eq:saddle_point_eqn}, and neglecting lubrication forces, we invert
\begin{equation} \label{inv_saddle_point_eqn}
    \begin{bmatrix}
        \mathcal{M}^{\dag} & \mathcal{B} \\
        \mathcal{B}^{T} & \bm{0}
    \end{bmatrix}^{-1}
    \begin{bmatrix}
        \begin{pmatrix}
            \hat{\mathbf{U}} \\
            \bm{0}
        \end{pmatrix} \\
        \bm{0}
    \end{bmatrix}
    =
     \begin{bmatrix}
        \mathcal{F} \\
        \hat{\bm{\Omega}}
    \end{bmatrix}.
\end{equation}
We are interested in the bottom set of equations that gives the angular velocities.
After block-inverting, we obtain
\begin{equation}
    \left( \mathcal{B}^{T} \cdot (\mathcal{M}^{\dag})^{-1} \cdot\mathcal{B} \right)^{-1} \cdot \mathcal{B}^{T} \cdot (\mathcal{M}^{\dag})^{-1} \cdot
    \begin{pmatrix}
        \hat{\mathbf{U}} \\
        \bm{0}
    \end{pmatrix} \\
    = \hat{\bm{\Omega}} .
\end{equation}
Only the first $3N$ columns of this matrix are of interest, yielding $\mathcal{A} \cdot \hat{\mathbf{U}} = \hat{\bm{\Omega}}$. 
From the perspective of a resistance formulation, $\mathcal{A} = -(\textbf{R}^{\text{L}\Omega})^{-1} \cdot \textbf{R}^{\text{LU}}$. 
We specify a nonzero $\bm{\Omega}_2$ with $\bm{U}_{2} = \bm{0}$ and construct the saddle point system, 
\begin{equation} \label{inv_saddle_point_no_constraint}
    \begin{bmatrix}
        \mathcal{A} & \mathcal{P}_{1} \\
        \mathcal{P}_{2}^{T} & \bm{0}
    \end{bmatrix}
    \begin{bmatrix}
        \hat{\mathbf{U}} \\
        \bm{\Omega}_{1}
    \end{bmatrix}
    = 
    \begin{bmatrix}
        \begin{pmatrix}
            \bm{0} \\
            \bm{\Omega}_{2}
        \end{pmatrix} \\
        \bm{0}
    \end{bmatrix}. 
\end{equation}
The projection matrices, $\mathcal{P}_{1}$ and $\mathcal{P}_{2}$,  are defined such that $\mathcal{P}_1 \cdot \bm{\Omega}_1 = \left( -\bm{\Omega}_1 \ \ \bm{0} \right)^{T}$ and $\mathcal{P}_{2}^T \cdot \hat{\mathbf{U}} = \bm{U}_2$.

Still, we need to constrain the moving particle to the desired shape of its trajectory.
For example, we can define the shape function, $g(\bm{x}_1(t)) = (\bm{x}_1 \cdot \bm{e}_{x})^2 + (\bm{x}_1 \cdot \bm{e}_{y})^2 - R^2$, and impose that the particle position lies along the manifold $g(\bm{x}_1) = 0$, a circular trajectory of radius $R$.
We obtain a constraint on the velocities by differentiating with respect to time, $\skew{4}{\dot}{g} = \bm{J}^{T} \cdot \bm{U}_{1}$.
Here, $\bm{J} = \nabla g|_{\bm{x}_{1}(t)}$ is the Jacobian vector, which is not to be confused with the Oseen tensor in the main text. 
We solve for the velocities so that $\skew{4}{\dot}g =0$, specifying that they are directed orthogonally to the constraining surface.
Using a Lagrange multiplier, $\lambda$, and augmenting the present saddle point problem,
\begin{equation}
    \begin{bmatrix}
        \mathcal{A} & \mathcal{P}_{1} & \mathcal{L}\cdot\bm{J} \\
        \mathcal{P}_{2}^{T} & \bm{0} & \bm{0} \\
        \bm{J}^{T}\cdot\mathcal{L}^{T} & \bm{0} & 0
    \end{bmatrix}
    \begin{bmatrix}
        \hat{\mathbf{U}} \\
        \bm{\Omega}_1 \\
        \lambda
    \end{bmatrix}
    = 
    \begin{bmatrix}
        \begin{pmatrix}
            \bm{0} \\
            \bm{\Omega}_{2}
        \end{pmatrix} \\
        \bm{0} \\
        0
    \end{bmatrix} .
\end{equation} 
The projection matrix, $\mathcal{L}$, maps the Jacobian onto the set of all particle translational velocities. The solution to this system of equations gives $\skew{3}{\dot}{\bm{x}}_1 = \bm{U}_{1}$, such that $\bm{U}_2 = \bm{0}$.  
Incorporating near-field lubrication is straightforward. 
However, for the trajectories we have imposed in this work, the multipole expansion plenty suffices. 

Time-integrating the particle velocities yields the desired trajectory, $\bm{x}_1(t)$, that induces the specified $\bm{\Omega}_{2}$.
We add a ``post-stabilization'' correction, $\Delta\bm{x}_{1}^\dag$, to the integrated position of the moving particle, following a two-step procedure,
\begin{eqnarray}
    \bm{x}^{*}_1(t_{n+1}) = \bm{x}_1(t_n) + \int_{t_n}^{t_{n+1}}\bm{U}_1 \text{d}t, \\
    \bm{x}_1(t_{n+1}) = \bm{x}^{*}_1(t_{n+1}) + \Delta\bm{x}_{1}^\dag,
\end{eqnarray}
to impose the shape constraint on its trajectory \cite{Cline2004-qr, Ascher1998-ll}.
We derive the displacement correction using a Newton approximation around the uncorrected position, $\bm{x}^{*}_1$, in the shape function,
\begin{equation}
    g(\bm{x}^*_1 + \Delta\bm{x}_{1}^\dag) \approx g(\bm{x}^*_{1})  + \bm{J}^{T}\cdot \Delta\bm{x}_{1}^\dag =0,
\end{equation}
provided the deviation of $g(\bm{x}^*_1)$ from zero is small.
The displacement-correction is obtained from the psuedo-inverse of the Jacobian, 
\begin{equation}\label{Newton_pstab_corr}
    \Delta \bm{x}_{1}^\dag = -\left( \bm{J}^{T} \cdot \left(\bm{J}\bm{J}^{T}\right)^{-1}\right)g(\bm{x}^*_{1}).
\end{equation}
We can improve the accuracy of our correction by extending the expansion to second-order, also known as Haley's method, 
\begin{equation} \label{2nd_order_expansion}
    \begin{split}
        g(\bm{x}^*_{1} + \Delta\bm{x}_{1}^\dag) &\approx g(\bm{x}^*_{1}) + \bm{J}^{T}\cdot\Delta \bm{x}_{1}^\dag + \bm{H}:\Delta\bm{x}_{1}^\dag\Delta\bm{x}_{1}^\dag, \\
        &\approx g(\bm{x}^*_{1}) + 
        \bm{G}^{T} \cdot \Delta\bm{x}_{1}^\dag = 0.
    \end{split}
\end{equation} 
We define the grouping, $\bm{G}^{T} = \bm{J}^{T} + \bm{H}\cdot\Delta\bm{x}^{\dag,0}_{1}$, in which $\bm{H} = \nabla \nabla g|_{\bm{x}^*_1(t)}$ is the Hessian matrix.
We approximately solve Eqn.~\eqref{2nd_order_expansion} using the psuedo-inverse of $\bm{G}^{T}$, letting the displacement, $\Delta\bm{x}^{\dag,0}_{1}$, be the correction from Newton's method, Eqn.~\eqref{Newton_pstab_corr}.

\subsection{Supplemental Movies}
\label{movies}

\textbf{Supplemental Movie 1:} This movie demonstrates the time-dependent rotational Brownian motion of a hemispherical fluorescent PMMA particle under very weak optical trap stiffness. The scale bar represents 2.5 \textmu m. The red circle indicates the orientation point of the particle, while the red cross marks the particle's center. Using these markers, the orientation vector, \(\bm{q}\), was determined.

\textbf{Supplemental Movie 2:} Rotation of the probing particle in a two-body configuration at a constant pair separation distance (r = 9 \textmu m), influenced by changes in the imposed velocity of the moving particle. The imposed velocities are (A) 35 \textmu m/s, (B) 60 \textmu m/s, (C) 100 \textmu m/s, and (D) 200 \textmu m/s. The scale bar represents 5 \textmu m.

\textbf{Supplemental Movie 3:} Rotation of the probing particle in a two-body configuration at a constant moving particle velocity (U = 200 \textmu m/s), influenced by changes in the pair separation. The pair separations are (A) 6 \textmu m (B) 7 \textmu m, (C) 9 \textmu m, (D) 11 \textmu m, (E) 15 \textmu m, (F) 20 \textmu m, and (G) 30 \textmu m. The scale bar represents 5 \textmu m.

\textbf{Supplemental Movie 4:} Rotation of the probing particle (particle 2) in a three-body configuration, influenced by the presence or absence of the third particle (particle 3). The video was captured with the imposed velocity of the moving particle (particle 1) set to 100 \textmu m/s and the distance between particle 2 and particle 3 maintained at 6 \textmu m. The minimum distance between particle 1 and particle 2 was set at 7 \textmu m. (A) Case without particle 3, (B) Case with particle 3. The scale bar represents 5 \textmu m.

\textbf{Supplemental Movie 5:} Rotation of the probing particle in a hexagonally-close-packed (HCP) lattice at various local area fractions. The HCP area fractions ($\phi_A$) are (A) $0.4627$, (B) $0.3543$, (C) $0.1874$, and (D) $0.1008$. The scale bar represents 5 \textmu m.

\textbf{Supplemental Movie 6:} Rotation of the probing particle with different trajectories of the moving particle. (A) Cassini oval trajectory, (B) Elliptical trajectory. The scale bar represents 5 \textmu m.

\bibliography{apssamp_old}

\end{document}